\documentclass[aps,prc,preprint,superscriptaddress]{revtex4-2}

\usepackage{graphicx}
\usepackage{dcolumn}
\usepackage{bm}
\usepackage{threeparttable}
\usepackage{amssymb}
\usepackage{amsthm}
\usepackage{amsmath}
\usepackage{bm}
\usepackage{lineno}
\usepackage{hyperref}
\usepackage{adjustbox}
\usepackage{xcolor}
\usepackage{multirow}
\usepackage[normalem]{ulem}

\begin{document}


\title{Collinear laser spectroscopy on neutron-rich actinium isotopes}



\author{Ruohong Li}
\email{ruohong@triumf.ca}
\affiliation{TRIUMF Canada's particle accelerator laboratory, Vancouver, BC, Canada, V6T 2A3}
\affiliation{Department of Physics, University of Windsor, Windsor, ON, Canada, N9B 3P4}
\affiliation{Department of Astronomy and Physics, Saint Mary's University, Halifax, NS, Canada, B3H 3C3}

\author{Andrea Teigelh\"ofer}
\affiliation{TRIUMF Canada's particle accelerator laboratory, Vancouver, BC, Canada, V6T 2A3}%

\author{Jiguang Li}
\email{li\_jiguang@iapcm.ac.cn}
\affiliation{Institute of Applied Physics and Computational Mathematics, Beijing 100088, China}%

\author{Jacek Biero{\'n}}	
\affiliation{Instytut Fizyki Teoretycznej, Uniwersytet Jagiello{\'n}ski, Krak{\'o}w, Poland}%

\author{Andr\'{a}s G\'{a}csbaranyi}
\thanks{current address: Van der Waals-Zeeman Institute, University of Amsterdam, Amsterdam, Netherlands}
\affiliation{TRIUMF Canada's particle accelerator laboratory, Vancouver, BC, Canada, V6T 2A3}%

\author{Jake Johnson}
\affiliation{Instituut voor Kern- en Stralingsfysica, KU Leuven, 3001 Leuven, Belgium}%

\author{Per J\"{o}nsson}
\affiliation{School of Technology, Malm\"{o} University, S-205 06 Malm\"{o}, Sweden}%

\author{Victoria Karner}
\affiliation{TRIUMF Canada's particle accelerator laboratory, Vancouver, BC, Canada, V6T 2A3}%

\author{Mingxuan Ma}
\affiliation{Institute of Applied Physics and Computational Mathematics, Beijing 100088, China}

\author{Martin Radulov}
\affiliation{TRIUMF Canada's particle accelerator laboratory, Vancouver, BC, Canada, V6T 2A3}

\author{Mathias Roman}
\affiliation{TRIUMF Canada's particle accelerator laboratory, Vancouver, BC, Canada, V6T 2A3}

\author{Monika Stachura}
\thanks{co-senior author}
\affiliation{TRIUMF Canada's particle accelerator laboratory, Vancouver, BC, Canada, V6T 2A3}%

\author{Jens Lassen}
\thanks{co-senior author}
\affiliation{TRIUMF Canada's particle accelerator laboratory, Vancouver, BC, Canada, V6T 2A3}%
\affiliation{Department of Physics and Astronomy, University of Manitoba, Winnipeg, MB, Canada,  R3T 2N2}%
\affiliation{Department of Physics, Simon Fraser University, Burnaby, BC, Canada, V5A 1S6}%


\date{\today}

\begin{abstract}
High-resolution collinear laser spectroscopy of neutron-rich actinium has been performed at TRIUMF's isotope separator and accelerator facility ISAC. By probing the $7s^2~^1S_0$ $\rightarrow$ $6d7p~^1P_1$ ionic transition, the hyperfine structures and optical isotope shifts in $^{225, 226, 228, 229}\!$Ac$^+$ have been measured. This allows precise determinations of the changes in mean-square charge radii, magnetic dipole moments, and electric quadrupole moments of these actinium isotopes. The improved precision of charge radii and magnetic moments clears the ambiguity in the odd-even staggering from previous studies. The electric quadrupole moments of $^{225, 226, 228, 229}\!$Ac are determined for the first time.
\end{abstract}


\maketitle

\section{Introduction}\label{Introduction}

Octupole-deformation has been observed in nuclei beyond the $N$ = 126 shell closure and around $N~\approx$ 134 and $Z~\approx$ 88 where the Fermi surface lies between single-particle states with total angular and orbital momenta, $j$ and $l$, both differing by three \cite{Butler}. The octupole deformation breaks the reflection symmetry in the intrinsic frame. This enables the coexistence of two closely spaced bands with opposite parity in the nuclear structure, which are connected by strong dipole transitions. The nuclear octupole deformation and closely spaced parity doublets can significantly enhance the Schiff moment, making these nuclear systems promising candidates to investigate charge conjugation and parity (CP) violation and explore physics beyond the Standard Model~\cite{Spevak, Victor}. Examples include Ra-EDM (electric dipole moment) measurements on the pear-shaped $^{225}$Ra nucleus~\cite{Parker}, as well as experiments with radioactive molecules such as RaF, AcF, and ThF that benefit from both nuclear octupole deformation and strong effective internal electric fields\cite{RaF, AcF, Th229, Th_Kia}.

Also within this region inverse odd-even staggering (OES) of nuclear charge radii along the isotopic chain was experimentally observed to coincide with the reflection asymmetry, as in $_{85}$At~\cite{Barzakh}, $_{86}$Rn~\cite{Borchers}, $_{87}$Fr~\cite{Coc, Dzuba, Budincevic}, $_{88}$Ra~\cite{Wendt, Wansbeek, Lynch}, and $_{89}\!$Ac~\cite{Verstraelen}. In normal/non-inverse cases the odd-$N$ isotopes have comparatively smaller charge radii than the average of the two neighboring even-$N$ isotopes, which is interpreted in terms of pairing and quadrupole correlation \cite{Reehal}. The inverse OES and reflection asymmetry are suspected to be strongly correlated, yet the underlying connections remain unclear \cite{Leander, Iimura, Verstraelen}, requiring further experimental data for clarification. 

Laser spectroscopy is a powerful tool to study the evolution of nuclear size, shape, and structures along an isotopic chain by precisely measuring the corresponding changes in spectroscopic atomic properties caused by the interaction of valence electrons and nucleus, i.e., isotope shift (IS) and hyperfine (HF) splitting \cite{Yang}. Located right in the center of the aforementioned octupole deformation region, neutron-rich actinium isotopes have been experimentally investigated using in-source laser resonance ionization spectroscopy (RIS)\cite{Verstraelen, Ferrer}, which allowed the determination of the changes in mean-square charge radii and magnetic dipole moments of $^{225-229}\!$Ac. The technique is limited by the GHz Doppler broadening inside the hot-cavity source, which precluded electrical quadrupole moments from being extracted. Three different energy density functional calculations were compared with the experimental results, and the need to include forced reflection symmetry in the model is strongly hinted. However, due to the large measurement uncertainty in the charge radius of $^{228}\!$Ac, the boundary of the inverse OES region could not be definitively confirmed for a conclusive comparison with the theoretical model. To resolve the ambiguity in previous works and obtain additional information on quadrupole moments, Doppler-free laser spectroscopy is needed.

In this work, we employed fast-beam collinear laser spectroscopy (CLS) to investigate the ISs and HF structures of neutron-rich actinium isotopes. Fast-beam CLS has been used for decades to study nuclear structures of exotic isotopes \cite{Campbell}. Through kinematic compression in a fast beam, Doppler broadening is essentially eliminated and the spectroscopic resolution is significantly enhanced. CLS is one of the few techniques where enhanced selectivity is coupled with increased sensitivity. This allowed more precise determinations of the changes in mean-square charge radii and the magnetic dipole moments of $^{225, 226, 228, 229 }\!$Ac, along with additional insights into their electric quadrupole moments.

\section{Experimental Setup }\label{Setup}

The neutron-rich actinium isotopes were produced at TRIUMF's isotope separator and accelerator facility (ISAC), where a 14~$\mu$A 480~MeV proton beam impinged on a thick UCx target coupled to a Re-lined hot-cavity ionizer\cite{target}. The generated actinium isotopes were laser ionized, extracted, and accelerated to 30~keV, mass-filtered by the ISAC high-resolution mass separator, and then delivered to the collinear laser spectroscopy beamline for study \cite{Levy, Voss2016}. The hot-cavity ionizer was operated at the typical current 230~A, which provided a temperature around 2000~$^{\circ}$C at the laser ionization region. The yield of laser ionized $^{226}\!$Ac during the experiment was about 4.5$\times$10$^7$~s$^{-1}$. The isotope beam intensities at ISAC are available in the online yield database \cite{ISAC_yield}.

\begin{figure}[htbp]
	\centering
	\includegraphics[width=0.7\textwidth]{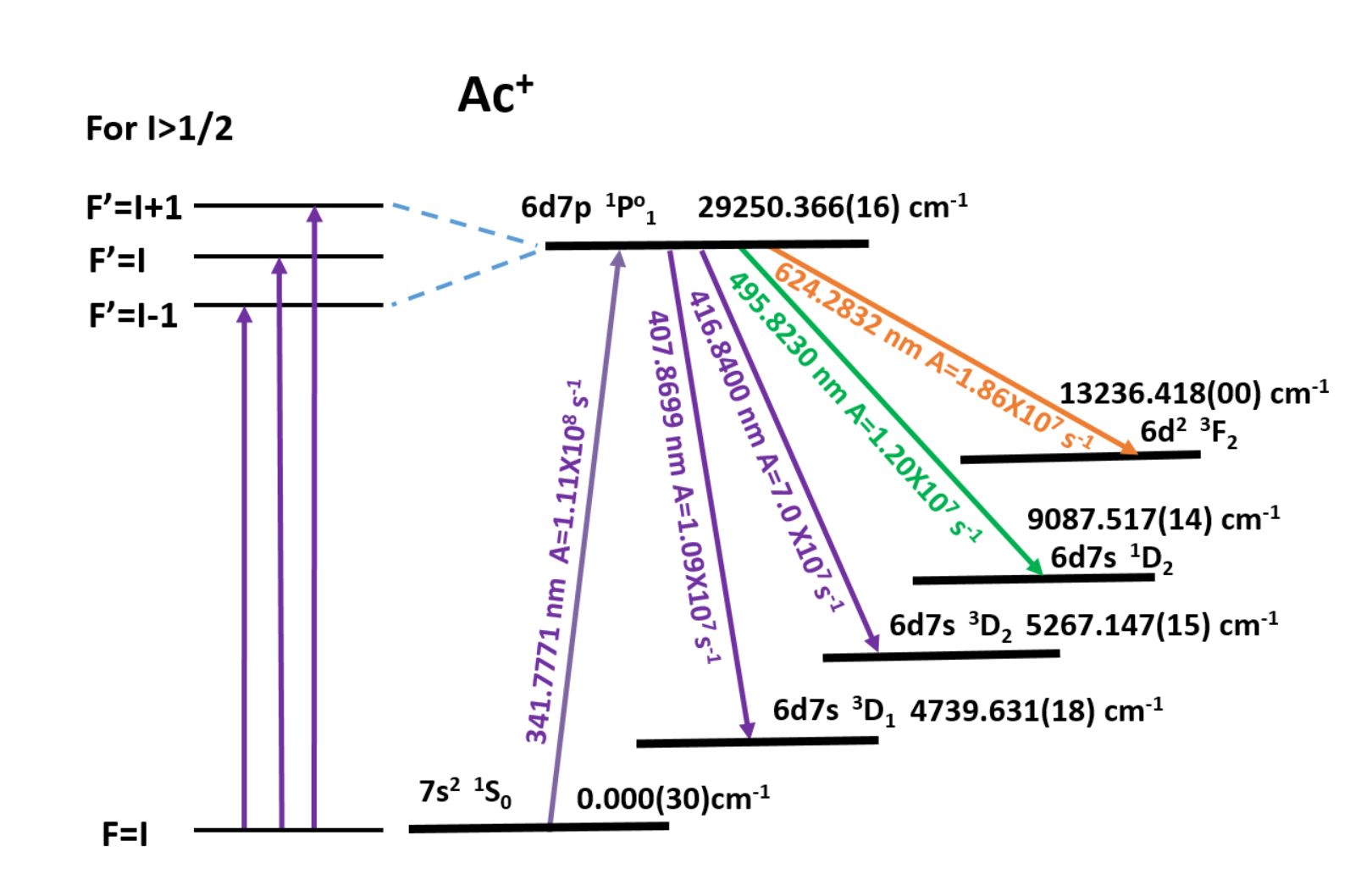}   
	\caption{Laser excitation scheme of Ac$^+$. The level energies, transition wavelengths, and transition probabilities are from the NIST database~\cite{NIST}, which has been recently updated on Ac. The wavelengths used are the observed wavelengths in air. }
	\label{fig:scheme}
\end{figure} 

Inside the polarizer beamline, the Ac$^+$ ion beam overlapped and interacted with a counter-propagating laser beam at 342 nm to excite the Ac$^+$ ions from the ground state to the upper state $6d7p~^1\!P_1$ of 29250.366~cm$^{-1}$ (Fig.~\ref{fig:scheme}). The potential of the fluorescent light collection region (LCR) was controlled by a scannable high voltage amplifier (Trek 609D-6), which allows to post-accelerate/decelerate the ion beam energy by up to \textpm 5~keV. During the data acquisition, the ion beam velocity was scanned by varying the voltage applied to the LCR, while the laser frequency remained constantly locked. This allowed the ions within the 50 mm-long LCR to be Doppler-tuned across the resonances. A spherical mirror light-collection system was used to efficiently collect fluorescence light from the excited state within 4$\pi$ solid angle. To reject scattered laser light from the excitation laser, a 10-nm bandwidth notch filter (Andover~415FS10-50) was used to selectively detect only the 417~nm transition photons from the upper level to the metastable state $6d7s~^3\!D_2$ of 5267.147~cm$^{-1}$. The collected fluorescence light was detected by a photon-counting photomultiplier (Hamamatsu H10682) and the output pulses were recorded using a digital multichannel-scaler data acquisition system~\cite{Voss2016}.

A single-mode ring dye laser (Coherent 899-21) was pumped by an 8.5~W frequency-doubled Nd:YAG laser (Spectra-Physics Millennia 10eV) to generate 270~mW fundamental laser at 684~nm, which was then frequency doubled into 1~mW 342~nm laser light via a BBO crystal in an external build-up cavity \mbox{(Spectra-Physics} WaveTrain). The 342~nm laser was transported over $\sim$17 meters with dielectric mirrors from the laser laboratory to the CLS beamline. After passing through the polarizer beamline the exit laser power was 0.4~mW. The laser beam was focused to $\sim$3~mm within the fluorescence LCR, matching the ion beam size. The dye laser was locked to a polarization-stabilized HeNe laser (Thorlabs HRS015B, wavelength accuracy of 10$^{-8}$) via a 300~MHz confocal Fabry-P\'{e}rot interferometer using the fringe off-set locking technique \cite{fringe}. The dye laser wavelength was also constantly monitored by a wavelength meter (HighFinesse WS7), which was routinely calibrated to the HeNe laser. During the experiment, the dye laser wavelength was measured as 14617.3755(1)~cm$^{-1}$ without running out of lock. The spectra were acquired as photon count rate as a function of post-acceleration voltage.   

\section{Results and discussion}\label{result}

\begin{figure*}[!h]
	\centering
	\includegraphics[width=1\textwidth]{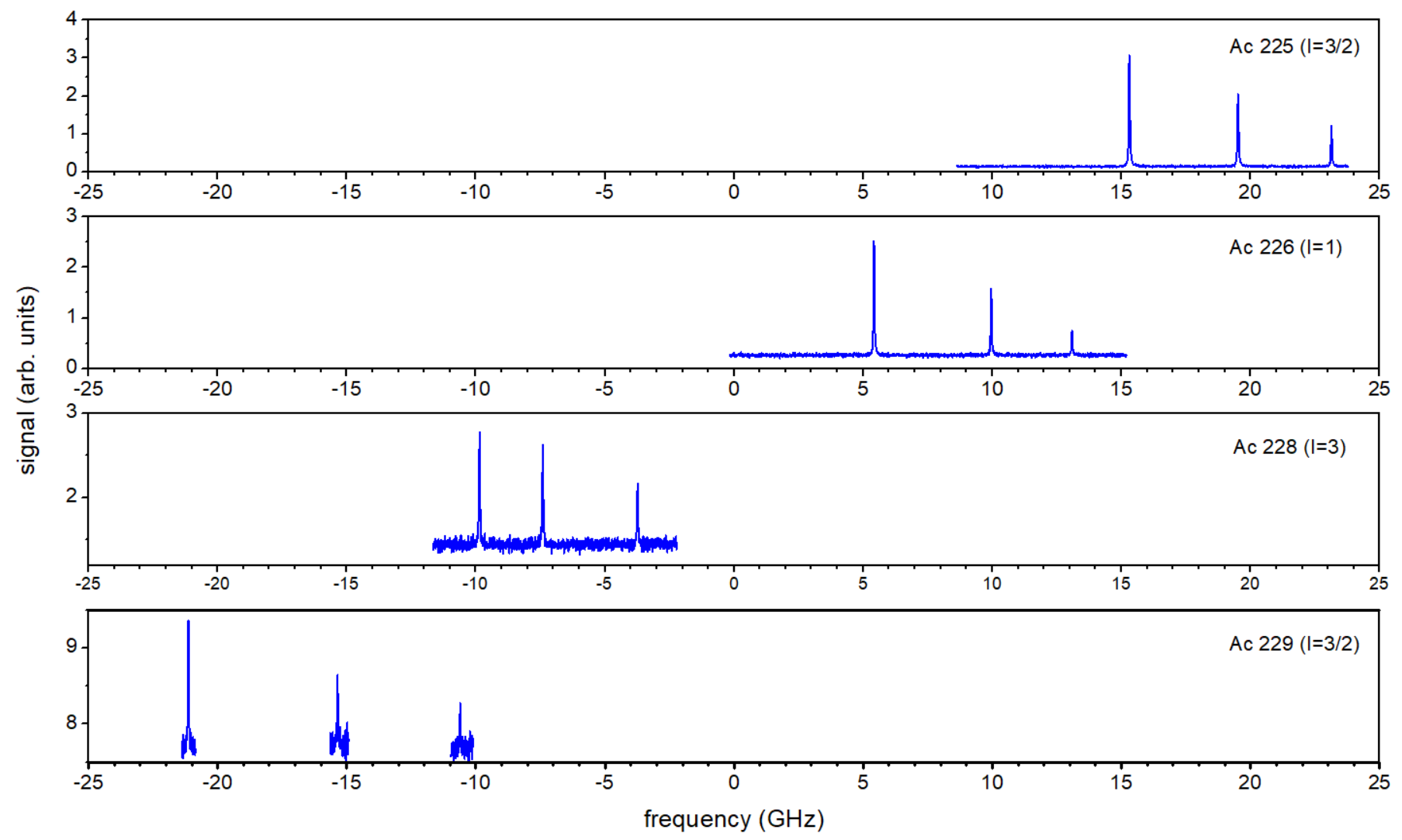}   
	\caption{HF spectra of $^{225, 226, 228, 229}\!$Ac$^+$, relative to the centroid frequency of the reference isotope $^{227}\!$Ac$^+$ \cite{NIST}.}
	\label{fig:spectra}
\end{figure*}

The observed high-resolution HF spectra of $^{225, 226, 228, 229}\!$Ac$^+$ are shown in Fig.~\ref{fig:spectra}. With an electronic angular momentum $J$~=~0 for the ground state, there is no HF splitting for the ionic ground state. The upper level, with $J$~=~1, splits into three HF states for any isotopes with $I$~$>$~1/2, which applies to all Ac isotopes investigated in this work. The relative amplitudes of HF transitions are constrained by angular momentum coupling coefficients: 
\begin{equation}\label{eq_A_HFS}
\begin{aligned}
A_{rel} (F \rightarrow F')~\propto~(2F+1)(2F'+1)\left\lbrace 
\begin{array}{ccc}
F' & F& 1 \\
J & J' & I
\end{array}
\right\rbrace ^2\\
\end{aligned}
\end{equation}
where $J$, $F$, and $J'$, $F'$ are the electronic and total angular momentum of the lower and upper level of the transition, respectively, and $I$ is the nuclear spin. Based on this formula three HF transitions for $^{225, 226, 228, 229}\!$Ac$^+$ should all have the relative amplitudes of 3:2:1.

The frequency shifts of the HF states relative to the original state prior to HF splitting (i.e., the centroid of the HF splittings) arise from the combined contributions of magnetic dipole and electric quadrupole interactions.:
\begin{equation}\label{eq_HFS}
\begin{aligned}
\delta\nu_{hfs} (F) = \dfrac{1}{2} AK + \dfrac{1}{2}B\dfrac{3K(K+1)-4I(I+1)J(J+1)}{2I(2I-1)2J(2J-1)}
\end{aligned}
\end{equation}
where $K$~=~${F(F+1)-I(I+1)-J(J+1)}$. Therefore, the frequency shifts of the HF transitions are given by $\delta\nu$$_{F \rightarrow F'}$ = $\delta\nu_{hfs} (F) - \delta\nu_{hfs} (F')$. Fig.~\ref{fig:spectra} shows the HF transition spectrum of each isotope, relative to the centroid frequency of the reference isotope $^{227}\!$Ac$^+$. The offset of the resonance centroid for each isotope from zero represents the isotope shift $\delta\nu$$^{A,227}$~=~$\nu$$^A$~-~$\nu$$^{227}$. In Fig.~\ref{fig:spectra} the scan voltage has been converted to frequency shift. Since there is no HF splitting in the ground state, no optical pumping between HF states is expected. Each HF spectrum was fitted using a $\chi^2$-minimization routine with the constraints of relative amplitudes 3:2:1 and identical linewidth for all peaks. The uncertainty of the photon counting data was assumed to be statistical as $\sigma$~=~$\sqrt{N}$. The fitting used the weight of $1/{\sigma}^2$. The reduced-$\chi^2$~($\chi_r^2$) ranged from 0.86 up to 1.26, which showed reasonable uncertainty estimation of the data. 

\begin{figure}[!h]
	\centering
	\includegraphics[width=0.7\textwidth]{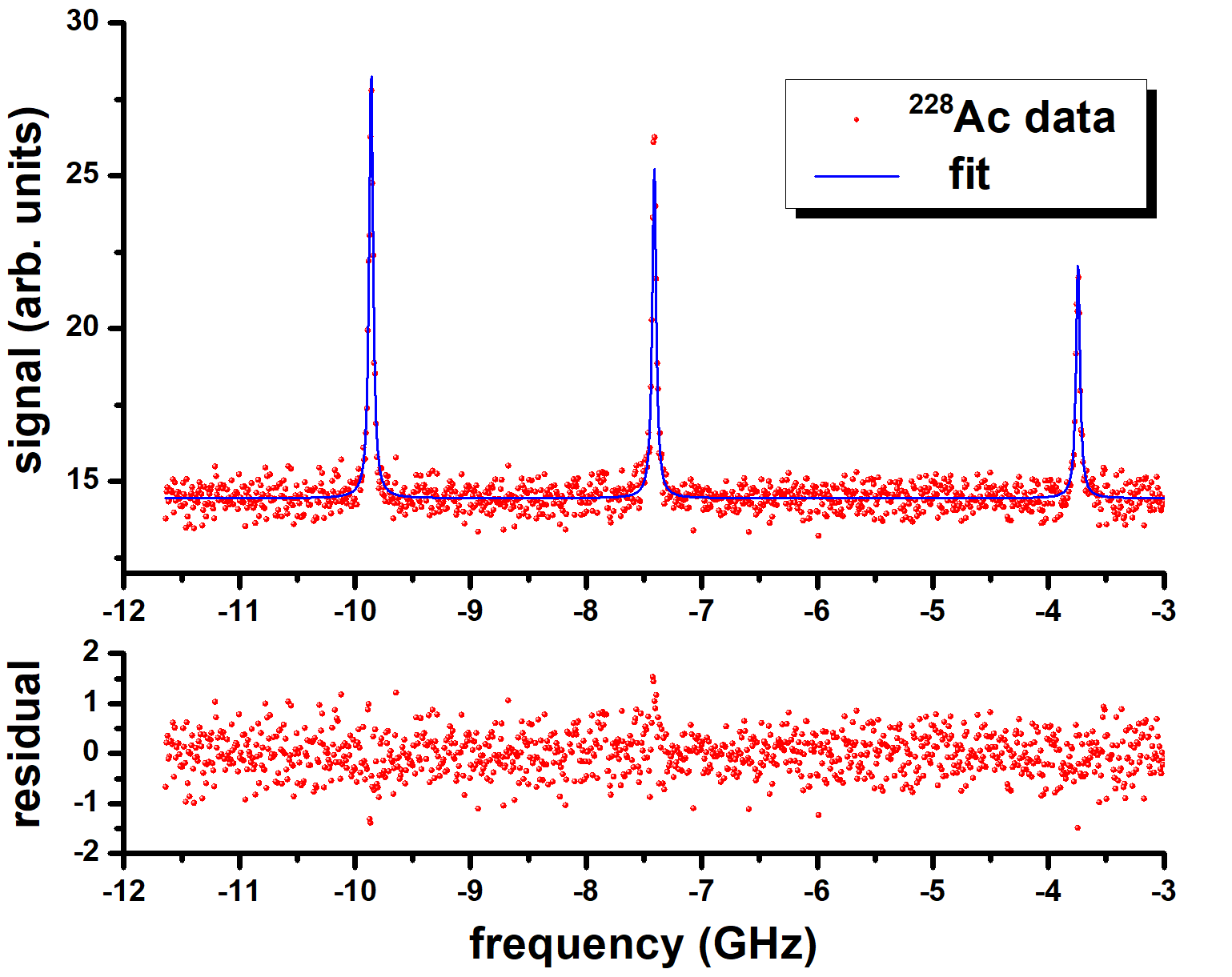}   
	\caption{HF spectrum of $^{228}\!$Ac$^+$, relative to the centroid frequency of the reference isotope $^{227}\!$Ac$^+$ \cite{NIST}. Lorentz profile was used and the relative intensity ratio of HF transitions was fixed. The resulting $\chi_r^2$ is 1.12 and the fitted linewidth is 41~MHz. }
	\label{fig:fit}
\end{figure}

Fig.~\ref{fig:fit} presents typical spectroscopy data and a fit of the HF spectrum of $^{228}\!$Ac$^+$ using Lorentz profile. The fit quality was tested with both pseudo-Voigt and Lorentz profiles. No significant difference was observed in the extracted resonance centroid when fitting with either profile, which implies the transition linewidth is likely determined by the natural linewidth of the upper level. The transition rate of the 342~nm transition is reported as $A_{ik}$~=~1.11$\times$10$^{8}$~s$^{-1}$ according to the NIST~database~\cite{NIST}. The recently updated transition rates for the other four transitions sharing the same upper level (see Fig.~\ref{fig:scheme}) contribute an additional total decay rate of 1.12$\times$10$^{8}$~s$^{-1}$ to the upper level. Therefore the upper level should have a natural linewidth of 35.4~MHz, which agrees with our observed linewidth of spectral peaks of 35-45~MHz. The Doppler linewidth of Ac$^+$ at a beam energy of 30 keV at ISAC is estimated to be 12 MHz. And the excitation laser linewidth is below 1 MHz. As a result, the observed spectrum linewidth should be dominated by the natural linewidth of the upper level as we observed.

The extracted HF constants $A$($^1\!P_1$), $B$($^1\!P_1$) and isotope shift $\delta\nu$$^{A,227}$ are presented in Table~\ref{table_HFS}. The fitting uncertainties (one standard deviation) are given in the brackets, which have been adjusted by $\chi_r^2$ (by multiplying by $\sqrt{\chi_r^2}$) when $\chi_r^2>1$ to give a conservative estimation of the errors, considering potential hidden uncertainties arising from the choice of fitting model and the underestimation of data uncertainty. For the isotopes with multiple scans/measurements, the weighted averages and their corresponding uncertainties are reported. The systematic uncertainty of the centroid frequency of the reference isotope $^{227}\!$Ac$^+$, arising from beam energy and wavelength determination (HighFinesse WS7 wavemeter), is estimated to be $\pm$30~MHz. This estimation is based on multiple measurements of the resonance frequencies of $^{8}$Li and $^{31}$Mg$^+$ conducted online from the same target station. The level energy of $6d7p~^1P_1$ for $^{227}\!$Ac$^+$ was listed as 29250.371~cm$^{-1}$ without clear uncertainty claim in NIST database \cite{NIST}, when we briefly reported the preliminary result of this experiment as a part of our facility updates \cite{polarizer}. Recently the Ac$^+$ data have been updated with clear level energy uncertainties and new information on transition probabilities. Fig.~\ref{fig:scheme} presents the updated atomic data. The level energy of $6d7p~^1P_1$ for $^{227}\!$Ac$^+$ is listed now as 29250.366(16)~cm$^{-1}$. The uncertainty of the 342~nm transition is 0.034~cm$^{-1}$ by combing 0.016~cm$^{-1}$ in quadrature with the ground-state uncertainty of 0.03~cm$^{-1}$. This will contribute an additional systematic uncertainty of 1.02~GHz to $\delta\nu$$^{A,227}$ values. This large systematic uncertainty could have been avoided by conducting CLS spectroscopy on $^{227}\!$Ac$^+$. However, $^{227}\!$Ac$^+$ was not included in this experiment due to radiation safety concerns. A recently installed alpha monitor within the beamline enables the measurement of internal contamination and will aid in obtaining safety approval for conducting CLS experiments on $^{227}$Ac$^+$ and other alpha-emitting species in the future.

\begin{table*}[h]
	\centering
	\caption{Summary of the half-lives, spins, and isotope shifts of the 342~nm transition of Ac$^+$ and the HF constants $A$ and $B$ for the level $6d7p$~$^1\!P_1$ for $^{225, 226, 228, 229}$Ac$^+$. Here the isotope shifts $\delta\nu$$^{A,227}$ are relative to $^{227}\!$Ac$^+$. The changes in mean-square charge radii $\delta \langle r^2 \rangle ^{A,227}$ are compared with the values measured previously via in-source spectroscopy on the atomic transition of 439~nm (in 2014 and 2016 \cite{Verstraelen}).}

	\begin{threeparttable}
		\begin{tabular}{>{\centering}p{0.04\textwidth} >{\centering}p{0.06\textwidth} >{\centering}p{0.17\textwidth} >{\centering}p{0.13\textwidth} >{\centering}p{0.13\textwidth} >{\centering}p{0.13\textwidth} p{0.01\textwidth} >{\centering}p{0.13\textwidth} >{\centering\arraybackslash}p{0.17\textwidth}}		
			\hline \hline           
            	&		&	\multicolumn{4}{c}{this work}	 &	&\multicolumn{2}{c}{Verstraelen's \cite{Verstraelen}}	\\
                \cline{3-6}  \cline{8-9}
			$A$	&	$I^{\pi}$	&	$\delta\nu$$^{A,227}$$\tnote{*}$	&	 $A$($^1\!P_1$)	&	$B$($^1\!P_1$)	&	$\delta \langle r^2 \rangle ^{A,227}$$\tnote{\textdagger}$ &	&\multicolumn{2}{c}{$\delta \langle r^2 \rangle ^{A,227}$$\tnote{\textdagger}$}  \\ 
			&			&	 (MHz)	&	 (MHz)	&	 (MHz)	&	(fm$^2$)&&	(fm$^2$)&	(fm$^2$)\\
			\cline{3-6}  \cline{8-9}																
			225	&		$\frac{3}{2}^-$ &	18012.80(11)	&	$-$1864.40(7)	&	364.41(14)	&	$-$0.128(8)[13]	&	&$-$0.206(3)[11]	&	$-$0.216(7)[11]	\\
			226	&		(1)$\tnote{\textborn}$	&		7783.04(21)	&	$-$2412.23(19)	&	190.32(21)	& 	$-$0.055(8)[6]	&&	$-$0.093(5)[5]	&		\\
			228	&	3$^+$	&	$-$7587.92(39)	&	$-$830.30(14)	&	871.49(74)	&	0.054(8)[5]	&&		&	0.095(27)[5]	\\
			229	&	($\frac{3}{2}^+$)$\tnote{\textborn}$	&	$-$17302.65(67)	&	$-$2524.58(47)	&	428.35(99)	&	0.123(8)[12]	&&		&	0.169(7)[9]	\\
			\hline 
			\hline 	
		\end{tabular}		
		\begin{tablenotes}
			\item[*] only presents the statistical uncertainty. For the total uncertainty, 1.05~GHz systematic error should be added, considering the uncertainty of the 342~nm transition energy of $^{227}\!$Ac$^+$ (1.02~GHz) plus the HV bias and WS7 wavemeter uncertainty (30~MHz).
			\item[\textdagger] the experimental uncertainty (including statistical and systematic) and theoretical uncertainty (originating from the atomic factors), are given in parentheses and brackets, respectively.				
			\item[\textborn] spin assignments are tentative.
		\end{tablenotes}
		\label{table_HFS}
	\end{threeparttable}
\end{table*}

\begin{figure}[htbp]
	\centering
	\includegraphics[width=0.6\textwidth]{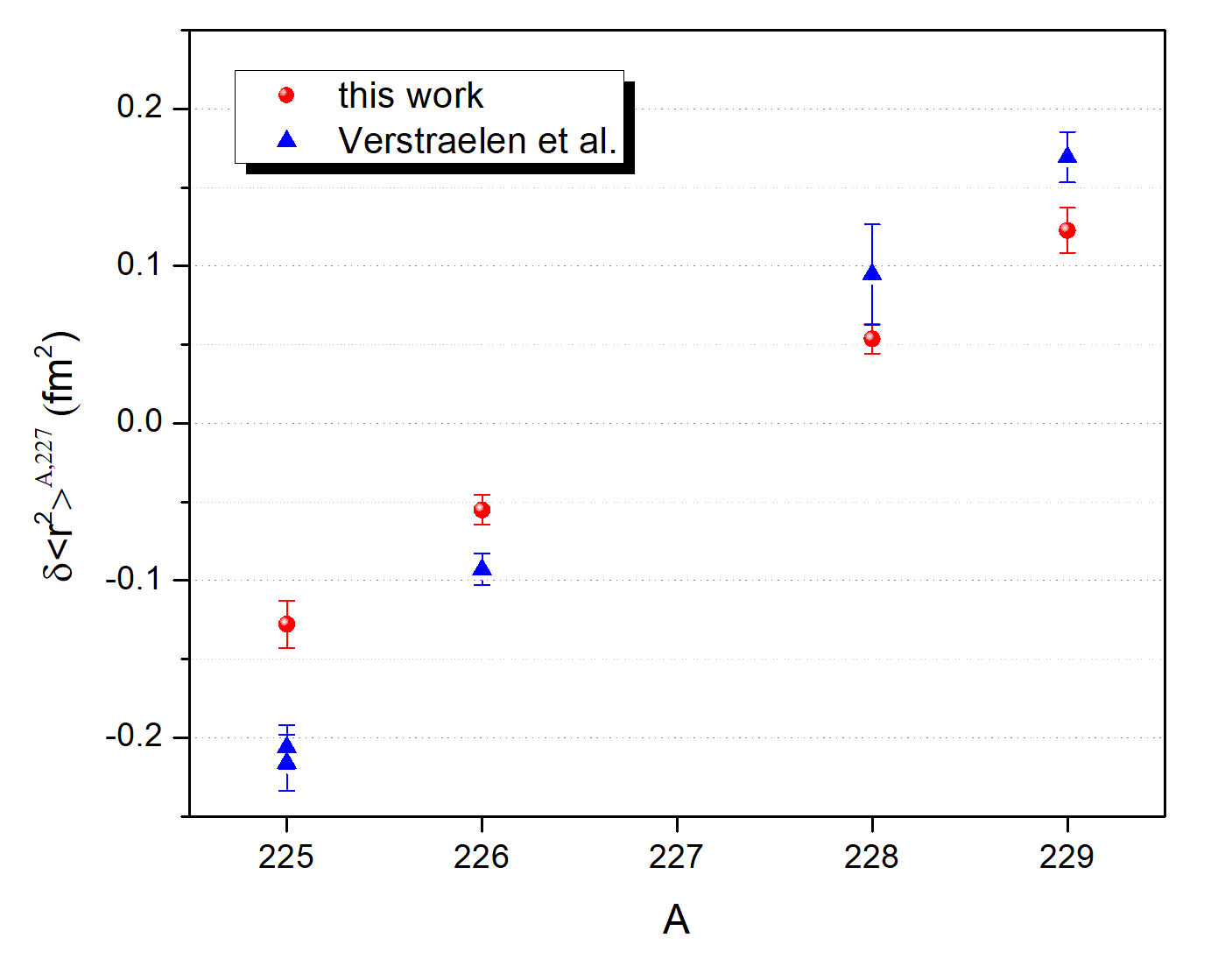}   
	\caption{The changes in mean-square charge radii $\delta \langle r^2 \rangle ^{A,227}$ obtained in this work, along with a comparison to the values of the previous work by Verstraelen et al.~\cite{Verstraelen}. The error bars include both experimental and theoretical uncertainties.}
	\label{fig:IS}
\end{figure} 

\subsection{Change in mean-square charge radii}\label{King-plot}
The isotope shift $\delta\nu$$^{A,A'}$ is related to the change in mean-square charge radius $\delta \langle$$r^2$$\rangle$$^{A,A'}$ as:
\begin{equation}\label{eq_IS}
\begin{aligned}
\delta \nu ^{A,A'} = M (\frac{A'-A}{A'A})+F \delta \langle r^2 \rangle ^{A,A'} 
\end{aligned}
\end{equation}
where $A'$ is the mass of the reference isotope. The atomic masses are taken from the AME2020 mass evaluation \cite{wangAME}. $M$ and $F$ are the mass and field isotope shift factors, respectively, specific to an atomic/ionic transition. Along the isotope chain of an element, adding more neutrons to the nucleus not only changes its mass - resulting in the mass isotope shift, but also alters its charge distribution - leading to the field isotope shift. The mass shift consists of normal mass shift and specific mass shift: 
$M~=~K_{\text{NMS}}+K_{\text{SMS}}$. 
For heavy elements like Ac, the mass-modification factor $g_m$~=~$\frac{A'-A}{A'A}$ is quite small, resulting in a rather minor contribution from the mass shift. Due to lack of the field and mass shift factors of the ionic transition used in this spectroscopic work, we carried out large-scale atomic-structure calculations by employing the GRASP package~\cite{GRASP2018, GraspManual2023} based on the multi-configuration Dirac-Hartree-Fock~(MCDHF) and the relativistic configuration interaction~(RCI) methods~\cite{Grant2007, MCDHFtheory}. The calculated factors for the ionic transition at 342~nm are $F_{342nm}~=~-138(14)$~GHz$\cdotp$fm$^{-2}$ and $M_{342nm}~=~9579(960)$~GHz$\cdotp$amu. The details of the theoretical calculation are presented in Sect.~\ref{MCDHF}.

Using these theoretical values, the change in mean-square charge radii $\delta \langle r^2 \rangle ^{A,227}$ were deduced and are presented in Tab.~\ref{table_HFS} and Fig.~\ref{fig:IS}, with comparison to the previous work of Verstraelen et al.~\cite{Verstraelen}. Understanding the systematic discrepancy in the comparison needs further development on atomic calculations or future independent experimental verification.  

\subsection {Theoretical calculations of atomic factors}\label{MCDHF}

The atomic-structure calculations consisted of two major phases. The first one was to generate a set of orbitals by using the MCDHF method~\cite{MCDHFtheory, GRASP2018, GraspManual2023}. The orbitals were divided into two categories. The first category comprises the occupied (also referred to as spectroscopic) orbitals, which are indeed occupied in the configuration(s) of interest. Spectroscopic orbitals must meet a number of specific requirements, related to their shapes, nodal structures, boundary, and orthonormality conditions. The second category includes correlation orbitals. These orbitals represent corrections to the wave function, in the sense that they account for various electron correlation effects. The requirements with respect to the shapes, nodal structures, and other properties of both orbital sets are described in detail in~\cite{GraspManual2023} and ~\cite{Bieron2009} 

Here we calculated the ground $7s^2~^1\!S_0$ and the excited $6d7p~^1\!P^o_1$ states for Ac$^{+}$ with a common core  $1s^22s^22p^63s^23p^63d^{10}4s^24p^64d^{10}4f^{14}5s^25p^65d^{10}6s^26p^6$. These two configurations were defined as the reference configurations for each symmetry. Also, test calculations revealed that the $7s7p$ configuration interacts strongly with the $6d7p$ configuration 
for $J^P=1^o$ symmetry, while $6d^2$ and $7p^2$ configurations interact with $7s^2$ for $J^P=0^e$, so these configurations were added to the sets of reference configurations with their respective symmetries. Except for the $6d$ and $7p$ subshells in the even-parity reference configurations, all occupied orbitals in the reference configurations were treated as spectroscopic ones. Spectroscopic orbitals were generated separately for each symmetry, with reference configurations described above, to minimize the average energy of the target atomic states (i.e.,~the EOL scheme~\cite{Grant1980, Grant1989}), using the self-consistent field~(SCF) method. For even parity the target atomic state was just the ground state, while for odd parity it included the five lowest states with $J^P=1^o$ symmetry, due to the significant correlation between $7s7p$ and $6d7p$.

Following that, correlation orbitals were generated, self-consistently optimized, and adapted to a specified multi-configuration expansion. The configuration expansions were formed in a single and restricted-double (SrD) approach, based on electron substitutions from the $n~\!\!\ge~\!\!4$ occupied orbitals to the set of correlation orbitals. The restriction was applied to all two-electron~(double) substitutions in such a way that only one electron from the core shells of the reference configuration ($4s^24p^64d^{10}4f^{14}5s^25p^65d^{10}6s^26p^66d7p$ for the odd parity, and $4s^24p^64d^{10}4f^{14}5s^25p^65d^{10}6s^26p^67s^2$ for the even parity) was allowed to be substituted for each particular configuration state function~(CSF). The set of the correlation orbitals was augmented in the layer-by-layer approach~\cite{GraspManual2023}, in which each subsequent layer was composed of orbitals with different angular symmetries, up to $g$. Seven layers of correlation orbitals were necessary to saturate the dominant electron correlation effects. While adding consecutive layers of correlation orbitals, we successively opened core shells, down to $n~\!\!=~\!\!4$.

In the layer-by-layer approach at each step only the last layer of correlation orbitals is variationally optimized; all layers generated in earlier steps (as well as all spectroscopic orbitals) in the MCDHF phase of our calculations were kept frozen. The correlation orbitals obtained with this model penetrated into the $n~\!\!\ge~\!\!4$ core region which facilitated capturing the correlation effects between electrons in these core shells in the later RCI computations. In addition, we adopted the Zero-First~(ZF) method~\cite{MCDHFtheory,GraspManual2023} in order to speed up the SCF procedures. 
The ZF method is almost equivalent to the second-order perturbation theory~\cite{Gustafsson2017,Dzuba2017c}. It should be emphasized that the orbital sets were generated independently for the ground state and for the excited states, in order to accurately describe their different electron correlations.  

In the second phase, we carried out the RCI computations to obtain the energy eigenvalues and the corresponding atomic state functions, by solving the eigenvalue equation of the atomic Hamiltonian~$\mathsf{H}$,
\begin{equation}\label{eq_H}
	(\mathsf{H} - E\mathbb{I}) \mathbf{X} = 0.
\end{equation}
Here, $\mathbb{I}$ is the unit matrix, and the vector $\mathbf{X}$ is a set of mixing coefficients, $\{c_1, c_2, \dots c_{N_{\text{CSF}}}\}$, corresponding to each CSF in the expansion of an atomic state function. In the atomic Hamiltonian, same as that used in the phase of generating the orbital basis, the nuclear potential was produced by the charge distribution in the form of two-parameter Fermi model~\cite{Parpia1992}
\begin{equation}\label{eq_rho}
	\rho(r) = \frac{\rho_0}{1+e^{(r-c)/a}},
\end{equation}
where $c$ is called the half-density radius (since $\rho(c) = \rho_0/2$), and $a$ is related to the nuclear skin thickness $t$ through the equation $a$~=~$t$/4ln$3$. Here $t$ is defined as a (relatively short) distance, in which the charge distribution falls from $\approx~$90\% to $\approx~$10\% of its central value.%

For Ac$^{+}$, with its 88 electrons, it is impossible to take into account the correlation effects between all electrons. To account for the dominant electron correlation effects, i.e.,~those involving the outer electronic shells, we generated the CSFs by employing the SrD substitutions from $n~\!\!\ge~\!\!4$ occupied orbitals in the reference configurations. The restriction imposed on the double substitutions was the same as that employed in the MCDHF phase of the calculations, but it was placed only on the $n~\!\!\le~\!\!5$ shells. In other words, the core-core electron correlation effects involving all $n~\!\!=~\!\!6$ shells were captured in the RCI phase of the calculations. Furthermore, the dominant triple and quadruple substitutions from the $n~\!\!\ge~\!\!6$ shells were also partly accounted for by employing the multi-reference~(MR) approach~\cite{Li2012, Filippin2017, Papoulia2021}. 
For this purpose, the sets of reference configurations were supplemented with $6s^26p^66d5f$, $6s^26p^67s8p$, $6s^26p^67p7d$, $6s^26p^67p8s$, $6s^26p^46d^37p$, $6s^26p^67d5f$, $6s^26p^56d^27s$, $6s^26p^56d7s^2$ for the excited state, and with $6s^26p^56d7s7p$, $6s^26p^46d7s7d$, $6s^26p^46d^27s^2$, $6s^26p^67s8s$, $6s^26p^47s^25f$, $6s^26p^56d7s5f$, $6s^26p^56d^27s7p$, $6s^26p^57s^25f7d$, $6s^26p^47s^27d^2$ for the ground state, respectively. 
To control the number of CSFs, only two layers of correlation orbitals were allowed to be active for the substitutions from the occupied orbitals with $n~\!\!\ge~\!\!6$.

The Breit interaction and one-electron quantum electrodynamics, i.e.~vacuum polarization and self-energy corrections~\cite{MCDHFtheory, GraspManual2023}, were also considered at the RCI phase. Moreover, relativistic mass-shift~(RMS) operators~\cite{Palmer1987, Shabaev1994, Tupitsyn2003, Gaidamauskas2011, Li2012a, Ekman2019}
\begin{equation}\label{eq_H_RMS}
	H_{\text{RMS}} = \frac{1}{2M} \sum_{ij} \left\{ \mathbf{p}_i \cdot \mathbf{p}_j - \frac{\alpha Z}{r_i} \left[ {\bm \alpha}_i + \frac{({\bm \alpha}_i \cdot \mathbf{r}_i)\mathbf{r}_i}{r_i^2} \right] \cdot \mathbf{p}_j \right\}
\end{equation}
were integrated into the RCI code of the {\sc grasp}2018 package for calculating mass shift factors~\cite{Yang2025}. The mass and field shift factors, $M$ and $F$, required for extracting the change in nuclear mean-square charge radii (see Eq.~\ref{eq_IS}) were obtained with the finite-field method from the energy eigenvalues~\cite{Korol2007, Zubova2016}. The resulting factors of the ionic transition at 342 nm are $F_{342nm}~=~-138(14)$~GHz$\cdotp$fm$^{-2}$ and $M_{342nm}~=~9579(960)$~GHz$\cdotp$amu.

\subsection{Odd-even staggering }\label{OES}
Quantifying the small-percentage odd-even staggering (OES), the staggering parameter, $\gamma_N$, is commonly used:
\begin{equation}\label{eq_gamma}
	\begin{aligned}
		\gamma_N = \dfrac{\delta\langle r^2\rangle^{N-1,N}}{\dfrac{1}{2}\delta\langle r^2\rangle^{N-1,N+1}}
	\end{aligned}
\end{equation} 

Here $N$ is conventionally set as an odd number. When $\gamma_N\!\!<$1, the nuclear chain exhibits normal OES; conversely, it displays inverse OES when $\gamma_N\!\!>$1. Employing $\gamma_N$ to visualize the OES trend eliminates the reliance on the atomic factor $F$, thereby avoiding the significant theoretical uncertainty associated with $F$. In addition, with the mass-modification factor $g_m$~=~$\frac{A'-A}{A'A}$~$\sim$~10$^{-5}$ for Ac, the uncertainty contribution from the mass shift to $\gamma_N$ can be neglected here as well.

\begin{figure}[!h]
	\centering
	\includegraphics[width=0.6\textwidth]{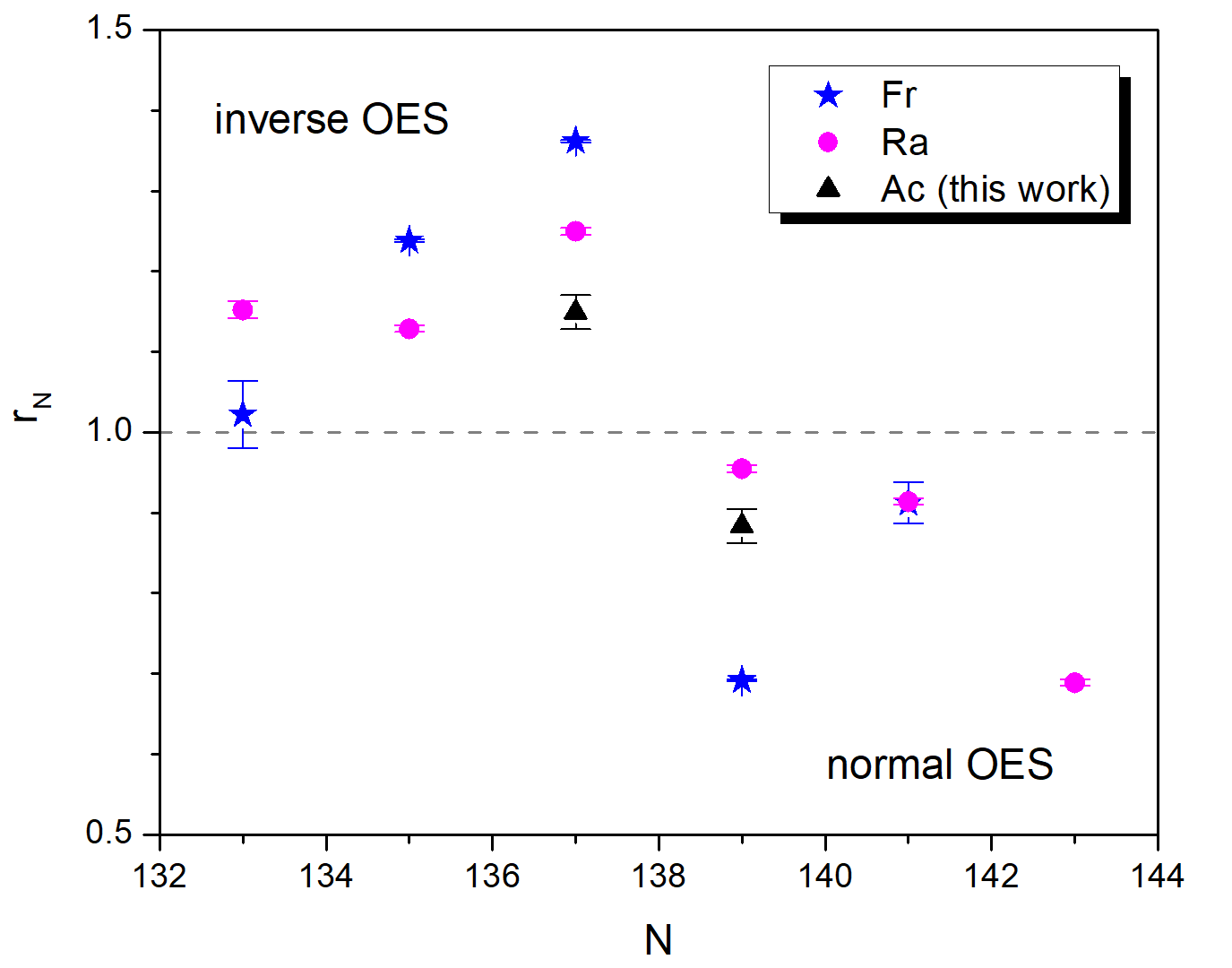}   
	\caption{Staggering parameter $\gamma_N$ of Ac (this work), Fr~\cite{Coc,Dzuba,Budincevic}, and Ra~\cite{Wendt, Wansbeek, Lynch}. A clear transition back to normal OES occurs at $N$~=~138 for Ac, consistent with the observations for Ra and Fr.}
	\label{fig:OES}
\end{figure}

The $\gamma_N$ values extracted from this work are shown in Fig.~\ref{fig:OES}. Inverse OES has been observed in a variety of isotopes around $N$ = 138~\cite{Barzakh, Borchers, Coc, Dzuba, Budincevic, Wendt, Wansbeek, Lynch}, including the previous work of Vertraelen~et~al.~\cite{Verstraelen} on actinium. However, the boundary of inverse OES for actinium could not be clearly defined due to the large uncertainty of $\delta \langle r^2 \rangle ^{228,227}$ in Vertraelen's work. In this work, with improved precision, a clear boundary returning to normal OES is observed at $N$~=~138 for Ac, similar to the cases of Ra and Fr (see Fig.~\ref{fig:OES}).

\subsection{Magnetic dipole moments and spectroscopic electric quadrupole moments}\label{dipole}
Based on the experimental values of the HF constants $A$ and $B$, the nuclear magnetic dipole moment $\mu$ and the spectroscopic electric quadrupole moment $Q_s$ can be extracted via:
\begin{equation}\label{eq_HFS_A_B}
\begin{aligned}
A = \dfrac{\mu B_{J}(0)}{IJ}, \;\;\;   B = e Q_{s}\langle V_{zz} \rangle,
\end{aligned}
\end{equation}
Eq.~\ref{eq_HFS_A_B} neglects HF structure anomaly caused by variations of the magnetization and charge distribution inside the nucleus of different isotopes, which is less than 1$\%$ in this region of the nuclear chart \cite{Verstraelen}.

\begin{figure}[!h] 
	\centering
	\includegraphics[width=1.0\textwidth]{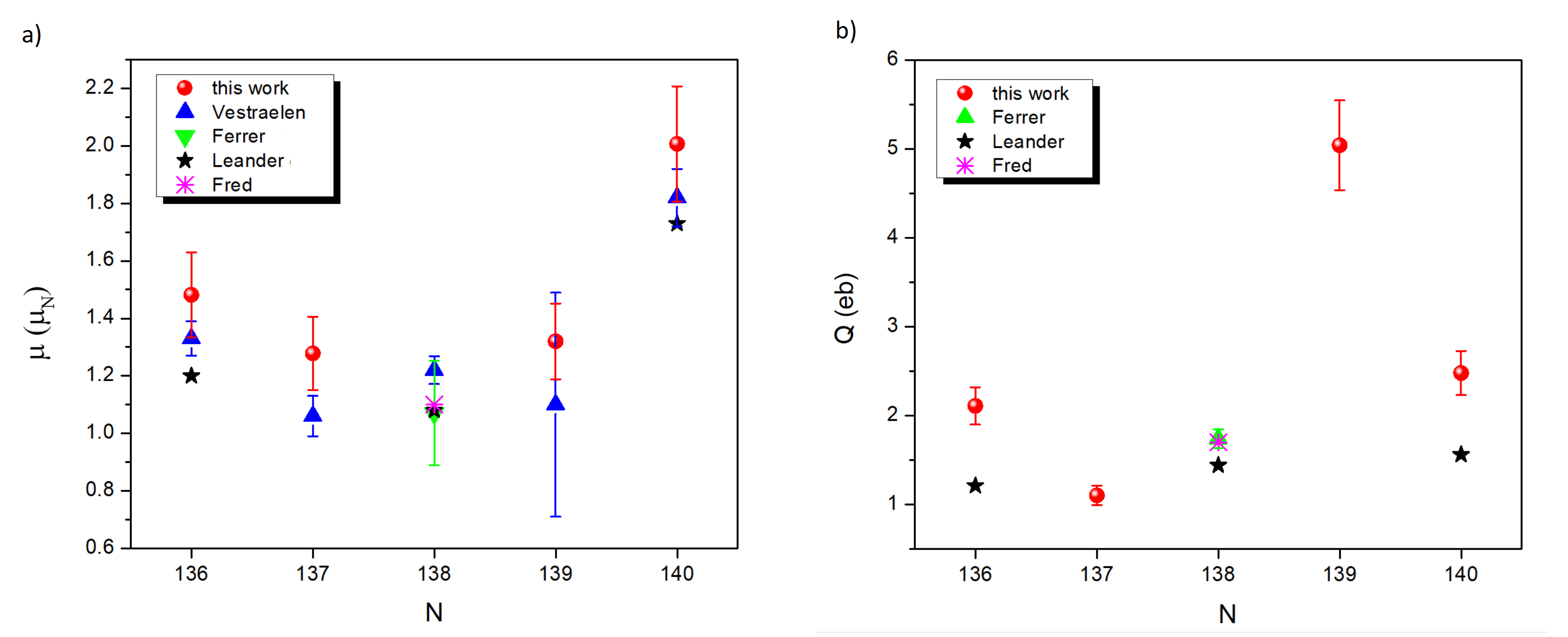}   
	\caption{Magnetic dipole moments $\mu$ and spectroscopic electric quadrupole moments $Q_s$ with comparisons to previous works: Verstraelen et al. (hot-cavity in-source spectroscopy + scaling from the $g$-factor obtained in $\gamma$-ray spectroscopy) \cite{Verstraelen}, Ferrer et al. (hot-cavity in-source spectroscopy + atomic theory) \cite{Ferrer}, Leander et al. (nuclear theory) \ cite{Leander}, Fred et al. (atomic spectroscopy + atomic theory in 1950s) \cite{Fred}.}
	\label{fig:u}
\end{figure}

\begin{table}[!ht]
	\begin{threeparttable}[]
	\centering
	\caption{Magnetic dipole moments $\mu$($\mu_N$) and quadrupole moments $Q_s$($eb$).}
	\renewcommand{\arraystretch}{0.48}		
		\bigskip
		\begin{tabular}{>{\centering}p{0.05\textwidth} >{\centering}p{0.05\textwidth} >{\centering}p{0.07\textwidth} >{\centering}p{0.15\textwidth} >{\centering}p{0.15\textwidth} >{\centering}p{0.02\textwidth} >{\centering}p{0.15\textwidth}>{\centering}p{0.12\textwidth}p{0.10\textwidth}p{0.08\textwidth}}		
				\hline \hline				
				& &	&\multicolumn{2}{c}{this work $\tnote{a}$}& &\multicolumn{4}{c}{$\mu$ $ $ (other works)}\\	
				\cline{4-5}\cline{7-10}                
				$A$	&	$N$	& I	&	 $A$($^1\!P_1$) 	&	$\mu$ 	& &	Verstraelen$\tnote{b}$	&	Ferrer$\tnote{c}$	&	Leander$\tnote{d}$ & Fred$\tnote{e}$	\\
				&	&	&	 (MHz)	&	 ($\mu$$_N$)	& &	 ($\mu_N$)	&	 ($\mu_N$)	&	 ($\mu_N$)	&	 ($\mu_N$)	\\
				\cline{4-5}\cline{7-10}															
				225	&	136	& 3/2$^-$&	$-$1864.40(7)	&	1.48203(6)	& &	1.330(19)\{40\}	&		&	1.20	&		\\
				226	&	137	& (1)&	$-$2412.23(19)	&	1.27834(10)	& &	1.06(4)\{3\} &		&		&		\\
				227	&	137	&3/2$^-$&		&		& &	1.220(18)\{30\}&	1.07[18]	&	1.08	&	1.1[1]	\\
				228	&	139	&3$^+$&	$-$830.30(14)	&	1.32002(22)	& &	1.10(9)\{30\}	&		&		&		\\
				229	&	140	&(3/2$^+$)&	$-$2524.58(47)	&	2.00682(38)	& &	1.82(3)\{7\}	&		&	1.73	&		\\
				\hline 
				\hline 	
		\end{tabular}		
		\bigskip
		\renewcommand{\arraystretch}{0.5}			
		\begin{tabular}{>{\centering}p{0.07\textwidth} >{\centering}p{0.07\textwidth} >{\centering}p{0.07\textwidth} >{\centering}p{0.15\textwidth} >{\centering}p{0.17\textwidth} >{\centering}p{0.02\textwidth} >{\centering}p{0.15\textwidth} >{\centering}p{0.15\textwidth} p{0.10\textwidth}}					
			\hline \hline
			&		&		&		\multicolumn{2}{c}{this work $\tnote{a}$}		&	&	\multicolumn{3}{c}{$Q_s$ (other works)}				\\
			\cline{4-5}	\cline{7-9}										
			
			$A$	&	$N$	&	$I$	&	 $B$($^1\!P_1$)	&	$Q_s$ &	&Ferrer$\tnote{c}$	&	Leander$\tnote{d}$	&	Fred$\tnote{e}$	\\
			&		&		&	 (MHz)	&	 ($eb$)	&	& ($eb$)	&	 ($eb$)	&	 ($eb$)	\\
			\cline{4-5}	\cline{7-9}														
			225	&	136	&   3/2$^-$	&	364.41(14)	&	2.1089(8)&	&		&	1.21	&		\\
			226	&	137	&	(1)	&	190.32(21)	&	1.1014(12)&	&		&		&		\\
			227	&	137	&	3/2$^-$	&		&	&	&	1.74[10]	&	1.44	&	1.7[2]	\\
			228	&	139	&	3$^+$	&	871.49(74)	&	5.0434(43)&	&		&		&		\\
			229	&	140	&	3/2$^+$	&	428.35(99)	&	2.4789(57)&	&		&	1.56	&		\\
			\hline 
			\hline 	
		\end{tabular}	
					
		\begin{tablenotes}
			\item[a] only statistical uncertainty is shown in the parentheses. The extra systematic uncertainty is 10$\%$ from the atomic theoretical calculation.
			\item[b] from Ref.~\cite{Verstraelen}. Statistical and systematic uncertainties (scaling from high-spin isomeric states of $^{215}\!$Ac $g$~=~0.920(13) \cite{Granados}) are given in parentheses and braces, respectively. 
			\item[c] from Ref.~\cite{Ferrer}. The total uncertainty is given in bracket, which is dominated by the theoretical calculation uncertainty of 17$\% $ and 6$\% $ for $\mu$ and $Q_s$, respectively. 
			\item[d] from Ref.~\cite{Leander}. Nuclear theoretical values using octupole-deformed Wood-Saxon shell model coupled to a reflection asymmetric rotor core. No error estimation was provided.
			\item[e] from Ref.~\cite{Fred}. The value was extracted from HF structure measurement and atomic calculations in the 1950s. The experimental uncertainty was claimed to be 10$\%$ (shown in the bracket), but the total uncertainty was hard to estimate.  
		\end{tablenotes}
		\label{table_u}
		
	\end{threeparttable}
\end{table}

$B_{J}(0)$ and $\langle V_{zz} \rangle$ are the magnetic field and the electric field gradient at the site of the nucleus, generated by the electrons, therefore they are strongly dependent on the electronic configuration and on the quantum numbers of the particular state. All previous HF structure measurements of Ac were studied by using the atomic transition at 439~nm from $6d7s^2$~$^3\!D_{3/2}$ to $6d7s7p$~$^4\!P_{5/2}$ \cite{Ferrer, Granados, Verstraelen}. Direct scaling using a known magnetic dipole moment is not feasible for analyzing the ionic transition data in this work. $B_{J}(0)$ and $\langle V_{zz} \rangle$ for the ionic state $6d7p$~$^1\!P_1$ were calculated by using the same computational model as that employed in Sect.~\ref{MCDHF} to determine the field and mass shift factors. However, the influence of the mass shift on the HF structure was not considered, meaning the atomic state function was not corrected by the mass shift. The obtained values are:
\begin{equation}\label{eq_atomic_factor}
	\begin{aligned}
	    B_{J}(0)_{342\text{nm}}=-1890(190)~\text{MHz} \,,\\
		\langle V_{zz} \rangle_{342\text{nm}} = -173(17)~\text{MHz} \,.\\
	\end{aligned}
\end{equation}
The extracted magnetic dipole moments $\mu$ and spectroscopic electric quadrupole moments $Q$ are presented in Tab.~\ref{table_u} and Fig.~\ref{fig:u} and compared to previous works, both experimental and theoretical. Since different atomic systems were used in these works, the total errors including the systematic errors from atomic calculation need to be considered for fair comparison. The details of the uncertainties of each work are shown in the caption of Tab.~\ref{table_u}. The majority of the total uncertainty stems from the theoretical calculations of atomic factors, which is typically $\sim$10$\%$. Verstraelen's results from in-source spectroscopy \cite{Verstraelen} are systematically higher than Leander's nuclear theory calculation~\cite{Leander} by about 0.1~$\mu_N$ (10$\%$), but lower than ours by 0.2~$\mu_N$ (20$\%$). 

It is noteworthy that Ferrer et al.~\cite{Ferrer} found a negative systematic offset 13$\%$ when comparing $\mu$ of $^{212-215}\!$Ac (the isotopes close to the magic number $N$~=~126) to shell model calculation. A similar shift was found when they compared the $g$-factor of $^{215}\!$Ac ground state with that of all h$_{9/2}$ based states in the $N$~=~126 isotones and that of high-spin isomeric states of the $^{215}\!$Ac extracted from $\gamma$-ray spectroscopy. Later Granados et al.~\cite{Granados} reanalyzed those data by scaling from the weighted average value $g$~=~0.920(13) of the two $^{215}\!$Ac isomers. The results then could well match large-scale nuclear shell-model calculations. This implied the 13$\%$ systematic discrepancy from the shell model arises from the theoretical atomic factors used. To avoid this problem, Verstraelen's $\mu$ values \cite{Verstraelen} were extracted by scaling from $^{215}\!$Ac data of Granados. The difference in the $\mu$ values of $^{227}$Ac ($N$ = 138 in Fig.~\ref{fig:u}a)) manifests methodological difference: Ferrer et al. used the atomic factors from MCDHF calculation, and Verstraelen et al. used the scaling method. This might not be an entirely fair statement since the experimental systems of Ferrer and Verstraelen are different, which could also introduce systematic uncertainty. Granados used the same experimental data as Ferrer and extracted $\mu$ of $^{227}\!$Ac to be 1.223(18)\{18\}~$\mu_N$, agreeing with Verstraelen's value of 1.220(18)\{30\}~$\mu_N$. It should be noted that, although 13$\%$ shows prominently as a systematic offset for multiple isotopes, it remains within the 17$\%$ error bar of the MCDHF calculation~\cite{Ferrer}. 

The systematic discrepancy between this work and Verstraelen's is in the opposite direction. Understanding this requires further investigation, either by improving theoretical calculations or through additional independent measurements of these Ac isotopes. Obtaining the HF spectrum of $^{227}$Ac$^+$ would provide a valuable reference isotope to bridge these experimental results, enabling a more meaningful comparison.

\subsection{\texorpdfstring{$g$-factor for even-$N$ actinium isotopes}{g-factor for even-N actinium isotopes}}\label{g_even}

\begin{figure}[!h]
	\centering
	\includegraphics[width=1\textwidth]{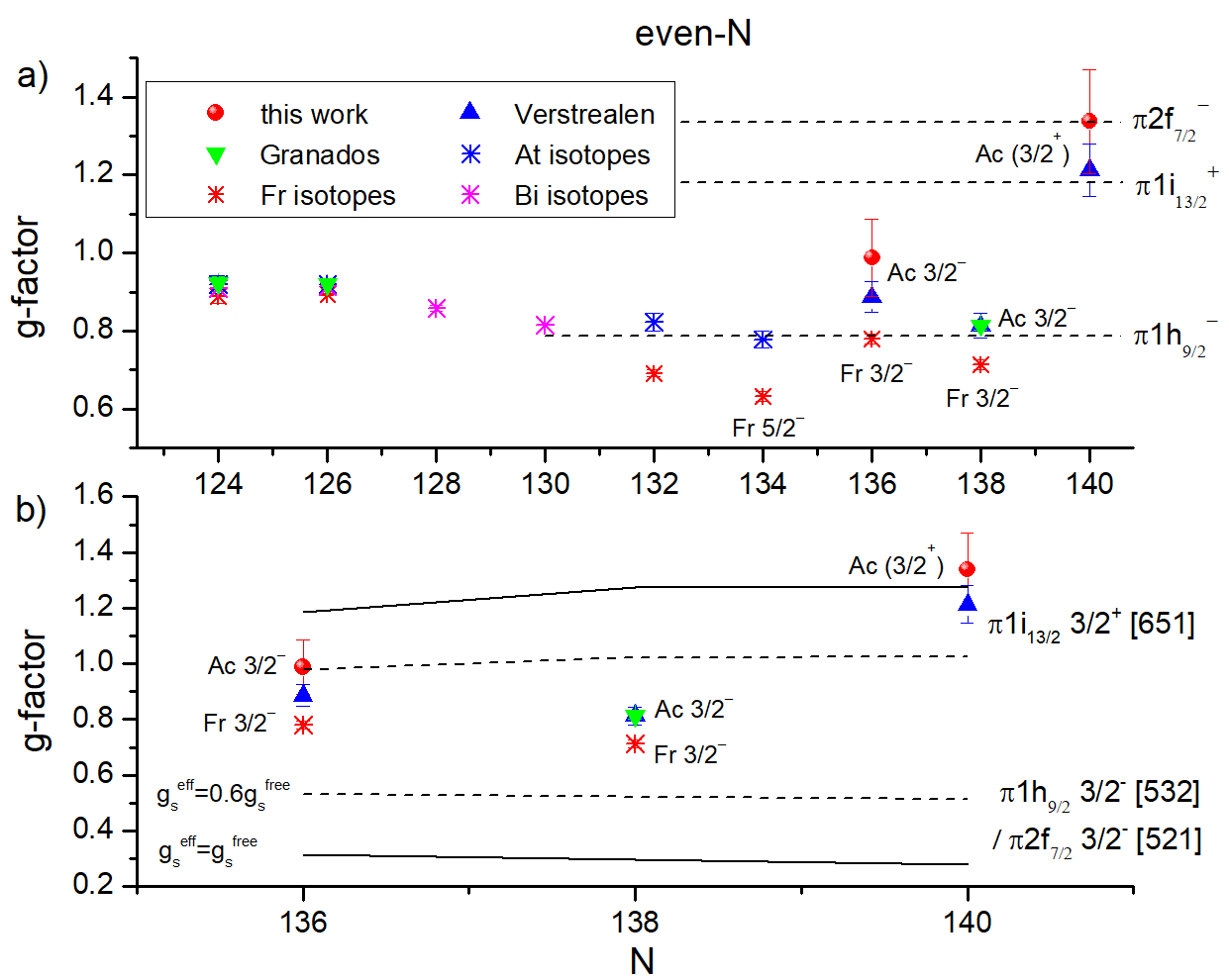}   
	\caption{Experimental $g$-factors for even-$N$ isotopes of At~\cite{Barzakh, Cubiss}, Fr~\cite{Budincevic, Coc}, Bi~\cite{Barzakh_Bi, Matt_Bi}, and Ac. Ac data include this and previous works. The previous works on Ac are: Verstraelen et al. (hot-cavity in-source spectroscopy + scaling from the $g$-factor obtained in $\gamma$-ray spectroscopy)~\cite{Verstraelen}, Granados et al. (gas-jet/hot-cavity in-source spectroscopy + scaling from the $g$-factor obtained in $\gamma$-ray spectroscopy)~\cite{Granados}.  
    \\
    a) The dashed line marks the $g_{sp}$ calculated using $g_s^\text{eff} = 0.6~g_s^\text{free}$. \\
    b) The dashed and solid line mark the $g$-factors of the Nilsson states calculated via Eq.~\ref{eq_g_odd_A} using $g_s^\text{eff}$ = $0.6~g_s^\text{free}$ and $g_s^\text{free}$, respectively. $\pi2f_{7/2}$ (3/2$^-$[521]) and $\pi1h_{9/2}$~(3/2$^-$[532]) have a same $g$-factor.}
	\label{fig:g1}
\end{figure}

The experimental nuclear $g$-factors have been calculated by $g\!=\!\mu/I$ (Fig.~\ref{fig:g1}). The \mbox{$g$-factors} of even-$N$ actinium ($Z$~=~89) isotopes are compared with the single-particle $g$-factor ($g_{sp}$) of the nearby proton orbitals $\pi1h_{9/2}$, $\pi1i_{13/2}$ and $\pi2f_{7/2}$. Another close-by orbital $\pi3s_{1/2}$ was ruled out since the spins of even-$N$ actinium isotopes studied here are all 3/2. The calculation employed the orbital $g$-factor ($g_l$) as 1 for a proton and the effective spin $g$-factor ($g_s^\text{eff}$) as 0.6~$g_s^\text{free}$, which is consistent with the methods in Ref.~\cite{Verstraelen, Budincevic}. Here $g_s^\text{free}$ is 5.587~\cite{CODATA} for a proton in a free particle state. The calculated values of $g_{sp}$ are 0.78 for $\pi1h_{9/2}$, 1.18 for $\pi1i_{13/2}$, and 1.34 for $\pi2f_{7/2}$, as shown by the dashed lines in Fig.~\ref{fig:g1}. 

The \mbox{$g$-factors} of the nearby even-$N$ isotones with odd-$Z$ ($_{83}$Bi~\cite{Barzakh_Bi, Matt_Bi}, $_{85}$At~\cite{Barzakh, Cubiss}, $_{87}$Fr~\cite{Budincevic, Coc}) are also plotted in Fig.~\ref{fig:g1} for comparison. At the neutron shell closure $N$~=~126, the {\mbox{$g$-factor}} values of these isotones overlap quite well. When $N$ increases, the \mbox{$g$-factor} gradually decreases with a constant slope~\cite{Barzakh} until $N$~=~134. The slopes are slightly different for different elements. From $N$~=~134 to 136, the \mbox{$g$-factor} of the Fr isotope chain jumps up, due to the spin change from 5/2 to 3/2. However, from $N$~=~136 to 138 it returns to a decreasing trend with a slope similar to that observed before the jump. Then from $N$~=~138 to 140, the \mbox{$g$-factor} dramatically increases from 0.78 of $\pi1h_{9/2}$ to 3.35 of $\pi3s_{1/2}$~\cite{Budincevic} (not shown in Fig.~\ref{fig:g1}, but in Fig.~7 in Ref.~\cite{Budincevic}), which implies the valence-proton state changes from $\pi1h_{9/2}$ to $\pi3s_{1/2}$. Namely, from $^{227}$Fr$_{140}$ upwards the nuclear ground state wave function starts to be dominated by a proton-intruder configuration, a proton hole in the $\pi3s_{1/2}$ orbital. Although this intrusion has not yet appeared in the isotone $^{229}\!$Ac$_{140}$, it likely occurs in $^{231}\!$Ac$_{142}$, where the assigned spin $I^\pi$ = 1/2$^{+}$ ~\cite{Thompson} matches with the intruder orbital. Due to the lack of data from $N$~=~128 to 134, it is unknown if there is a similar small jump around $N$~=~134 for Ac. In Fig.~\ref{fig:g1} we can see the dominant ground state for $^{225}\!$Ac$_{136}$ and $^{227}\!$Ac$_{138}$ is still $(\pi1h_{9/2})^-$. 

Despite the systematic shift observed in our data compared to previous studies on Ac~\cite{Verstraelen, Ferrer}, the sudden increase in $g$-factor at $N$~=~140 is consistent, which can be explained by a jump to a higher-$j$ proton configuration, such as $(\pi1i_{13/2})^+$ or $(\pi2f_{7/2})^-$, or an admixture thereof. The spin of $^{229}\!$Ac$_{140}$ has been tentatively assigned as 3/2$^+$~\cite{NNDC}. Assuming the assigned parity is correct, the proton configuration can only be $(\pi1i_{13/2})^+$, if reflection-symmetry holds. With octupole deformation present, it could be a mixture of two opposite-parity orbitals $(\pi1i_{13/2})^+$ and $(\pi2f_{7/2})^-$. As stated in the introduction, the correlation within two nearly degenerate single-particle orbitals with $\varDelta l$~=~3 and $\varDelta j$~=~3 plays a significant role~\cite{Arita} in the origin of ground-state octupole deformations. In this $Z$ region, it implies the strong coupling between $(\pi1i_{13/2})^+$ and $(\pi2f_{7/2})^-$ orbitals, which are nearly degenerate in the realistic nuclear mean field potential. 

Despite the well-known deformation in the region of $N\!>$132 with $\beta_2\!\approx$~0.16 \cite{Leander1984}),
the single-particle $g$-factors still can roughly match with the experimental data of Fr, At, and Ac, which gives guidance in assigning the ground-state orbital. Due to the deformation, the orbital level is no longer degenerate and Nilsson states should be more suitable to interpret the data. Using the Nilsson diagram, three possible states nearby are 
$\pi1h_{9/2}$~(3/2$^-$[532]), 
$\pi1i_{13/2}$~(3/2$^+$[651]), and 
$\pi2f_{7/2}$~(3/2$^-$[521]). 
The $g$-factor of a deformed odd-$A$ nucleus can be calculated using the method outlined in Ref.~\cite{Verstraelen} (and the references therein), along with the parameters employed:

\begin{equation}\label{eq_g_odd_A}
	\begin{aligned}
		g_{\text{odd-A}} = g_R+(g_K-g_R)\dfrac{K^2}{I(I+1)}(1+\delta_{K, 1/2}(2I+1)(-1)^{I+1/2}b_0)
	\end{aligned}
\end{equation} 

where 
\begin{equation}\label{eq_g_K}
	\begin{aligned}
		g_K = g_l+\frac{1}{K}(g_s-g_l)\langle s_z \rangle\\
	\end{aligned}
\end{equation} 

The collective rotational gyromagnetic ratio $g_R$ was taken as 0.4 for the single particle proton \cite{Lamm}. The angular momentum projection on the symmetry axis $K$ was assumed to be equal to $I$ for well-deformed nuclei. $\langle s_z \rangle$ were taken from Fig.~10 of Ref.~\cite{Leander1984}, which are the values under the constraint of reflection symmetry. The $g_s^\text{eff}$ value was set as $0.6~g_s^\text{free}$ and $g_s^\text{free}$. The calculated $g$-factors of these possible Nilsson states are shown with our experimental data in Fig.~\ref{fig:g1}. The $g$-factor for $\pi2f_{7/2}$ (3/2$^-$[521]) is the same as that of $\pi1h_{9/2}$~(3/2$^-$[532]).
  
Similar to the analysis in Verstraelen's work~\cite{Verstraelen}, the comparison reveals a mixing of Nilsson states with opposite parities $\pi1h_{9/2}$~(3/2$^-$[532]) (or $\pi2f_{7/2}$ (3/2$^-$[521])) and $\pi1i_{13/2}$~(3/2$^+$[651]) at $^{225}\!$Ac$_{136}$ and $^{227}\!$Ac$_{138}$, but a significant reduction in mixing at $^{229}\!$Ac$_{140}$. It confirms that the ground state of $^{229}\!$Ac$_{140}$ is dominated by $\pi1i_{13/2}$~(3/2$^+$[651]), implying no obvious octupole deformation. The conclusion aligns with the prominent observation of the parity doublets in $^{225, 227}\!$Ac but not obvious in $^{229}\!$Ac\cite{Leander}. The boundary of this change is consistent with the confirmed boundary of inverse OES at $N$~=~138 in Sect.~\ref{OES}.

\begin{figure}[!h]
	\centering
	\includegraphics[width=1\textwidth]{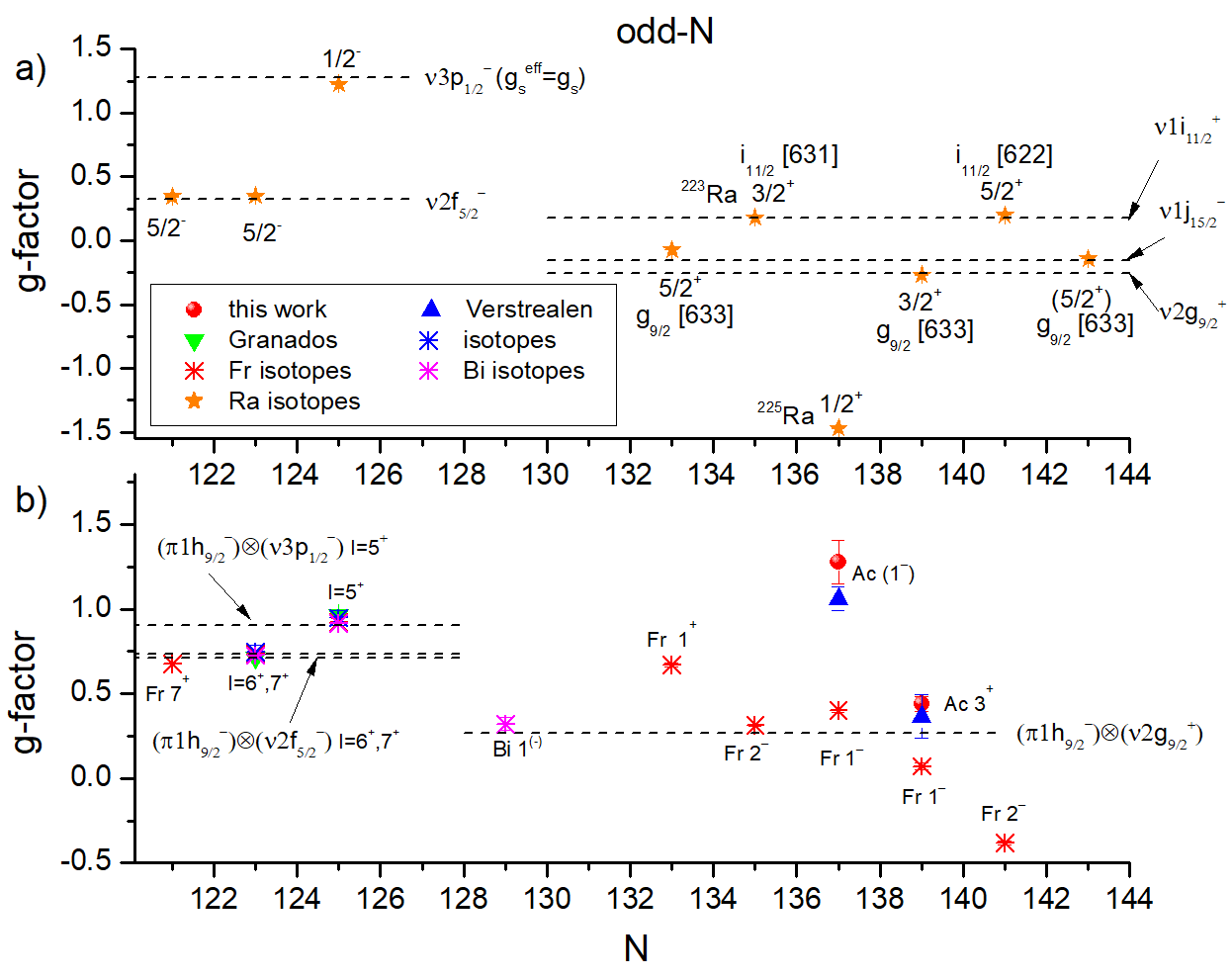}   
	\caption{Experimental $g$-factors for odd-$N$ isotopes of Ra~\cite{Wendt, Lynch}, At~\cite{Barzakh, Cubiss}, Fr~\cite{Budincevic, Coc}, Bi~\cite{Barzakh_Bi, Matt_Bi}, and Ac. Ac data include this and previous works. The previous works on Ac are: Verstraelen et al. (hot-cavity in-source spectroscopy + scaling from the $g$-factor obtained in $\gamma$-ray spectroscopy)~\cite{Verstraelen}, Granados et al. (gas-jet/hot-cavity in-source spectroscopy + scaling from the $g$-factor obtained in $\gamma$-ray spectroscopy)~\cite{Granados}.   
    \\
    a) even-$N$ Ra isotopes \cite{Wendt, Lynch}. The data points without labeled $I^\pi$ have the spin and parity of 9/2$^-$. The $g_{sp}$ values were calculated using $g_s^\text{eff} = 0.6~g_s^\text{free}$, except for the $\nu3p_{1/2}^-$ orbital, where $g_s^\text{eff} = g_s^\text{free}$ was used. \\ 
    b) The dashed marks the $g_\text{emp}$ values calculated using the additivity relation (Eq.~\ref{eq_g_emp}). At $N = 121-125$, a $g$-factor of 0.92 was used for the $\pi1h_{9/2}^-$ orbital. This value was estimated based on the average of the experimental data from Ac, Fr, At, and Bi, as shown in Fig.~\ref{fig:g1}.}
	\label{fig:g2}
\end{figure}

\subsection{\texorpdfstring{$g$-factor for odd-$N$ actinium isotopes}{g-factor for odd-N actinium isotopes}}\label{g_odd}

The odd-$N$ isotopes of actinium ($Z$~=~89) are odd-odd nuclei. Assuming single-particle coupling and using the additivity relation to understand the nucleon orbital occupations~\cite{Blin}, the empirical $g$-factor can be estimated as:
\begin{equation}\label{eq_g_emp}
	\begin{aligned}
		g_\text{emp} = \dfrac{1}{2}[(g_\pi +g_\nu)+(g_\pi-g_\nu)\dfrac{j_\pi(j_\pi+1)-j_\nu(j_\nu+1)}{I(I+1)}]
	\end{aligned}
\end{equation}
Here $I$ is the nuclear spin; $g_\pi$($j_\pi$) and $g_\nu$($j_\nu$) are the $g$-factor (angular momentum number) for proton($\pi$)  and neutron($\nu$), respectively. From the even-$N$ isotope analysis, the proton orbitals are $\pi1h_{9/2}$, $\pi1i_{13/2}$ and $\pi2f_{7/2}$. The nearby neutron orbitals are $\nu1i_{11/2}$ and $\nu1j_{15/2}$. Using the additivity relation, the $g$-factors of these nuclear configurations with the proton and neutron orbitals are shown in Fig.~\ref{fig:g2}a, with the experimental data of odd-$N$ Ra isotopes~\cite{Lynch}. The assigned neutron orbital of these Ra isotopes can be used to estimate the neutron orbital of the neighboring Ac isotopes.  

The additivity relation works well for the nuclei of At, Fr, Bi, and Ac near the magic number $N$~=~126. But when $N$ reaches 132 and beyond, the deformation starts to become significant, although Fr isotopes at $N$~=~135, 137, and 139 still show the rough agreement with the additivity value of $(\pi1h_{9/2}^-)\otimes(\nu2g_{9/2}^+)$.

For even-$N$ Ac isotopes with $N\!\!>$132 the coupling rule for deformed nuclei Eq.~\ref{eq_g_odd_A} was used. For odd-odd nuclei, the formula was modified to:
\begin{equation}\label{eq_g_odd_odd}
	\begin{aligned}
		g_{\text{odd-odd}} = \dfrac{K}{I(I+1)}(g_{K\pi}K_\pi + g_{K\nu}K_\nu+g_R\dfrac{I^2+I-K^2}{K})
	\end{aligned}
\end{equation} 
Here $g_K$ were calculated via Eq.~\ref{eq_g_K}, $g_R$~=~$Z/A$~$\approx$~0.4 was adopted for odd-odd nuclei in this region~\cite{Leander}, and $I$~\!\!=~\!\!$K$ was assumed. 
The Ra isotopes at $N$~=~133, 135, and 137 have been assigned ground-state spin as 5/2, 3/2, and 1/2, which align with the orbital-filling sequence at the $\beta_3$~=~0.1 in the single-particle plot of Fig.~4 for neutrons in Ref.~\cite{Leander}. The corresponding $\langle s_z \rangle$ values for neutron orbitals on the figure were adopted (listed in Tab.~\ref{table_odd}) to calculate $g_{K\nu}$ via Eq.~\ref{eq_g_K}. The sequence also matches with that in the $\epsilon_3$~=~0.08 plot in Fig.~4 of Ref.~\cite{Leander1984}, which used a folded Yukawa single-particle potential instead of the Woods-Saxon deformed shell model in Ref.~\cite{Leander}. Similarly, $\langle s_z \rangle$ for the proton orbitals were also obtained from Fig.~3 of Ref.~\cite{Leander} and Fig.~5 of ~\cite{Leander1984}. The $\langle s_z \rangle$ values from both theoretical models were found to be identical. 

Meanwhile, we can also use the $g$-factor of the neighboring odd-$N$ Ra isotope and that of even-$N$ Ac to estimate the $g_{K\nu}$ and $g_{K\pi}$, respectively, using Eq.~\ref{eq_g_odd_A}. Specifically, $g$~=~$-$1.468 for $^{225}\!$Ra \cite{Lynch} and $g$~=~0.988 for $^{225}\!$Ac (this work) were used to calculate the $g_{K\nu}$ and $g_{K\pi}$ for $^{226}\!$Ac; $g$~=~$-$0.269 for $^{227}$Ra \cite{Lynch} and $g$~=~0.813 ofr $^{227}\!$Ac~\cite{Verstraelen} were used to calculate the $g_{K\nu}$ and $g_{K\pi}$ for $^{228}\!$Ac. These half theoretical and half empirical results are listed as $g_\text{th-emp}$ in Tab.~\ref{table_odd}. More consideration was put for $K$~=~1/2 (see Eq.~\ref{eq_g_odd_A}, extra term with $b_0$), which applies to the valence neutron orbital of $^{226}\!$Ac (the $g_{K\nu}$ was assumed to be the same as that of the neighboring $^{225}\!$Ra). We adopted $b_0$~=~$-$1.64 as deduced in Ref.~\cite{Verstraelen}. The theoretical results are presented in Tab.~\ref{table_odd} and compared with the experimental ones, which show basic agreement. 

\begin{table*}[!h]
	\centering
	\caption{Comparison of experimental and theoretical values of $g$-factor}
	\setlength{\tabcolsep}{6pt} 
	\begin{threeparttable}
		\begin{tabular*}{\textwidth}{@{\extracolsep{\fill}}c c c c c c c}
			\hline \hline	
			\multirow{2}{*}{Isotopes} & \multirow{2}{*}{$I^\pi$} & ($\langle s_z \rangle_\pi$, $\langle s_z \rangle_\nu$) & \multicolumn{2}{c}{Theory} & \multicolumn{2}{c}{Experiment} \\
			\cline{4-5} \cline{6-7}
				&	&  \cite{Leander, Leander1984}	&	$g_\text{th}$$\tnote{a}$ & $g_\text{th-emp}$$\tnote{b}$ & This work & Verstraelen~\cite{Verstraelen} \\
			\hline															
			$^{226}\!$Ac & (1$^-$) & (0, 0.2) & 1.18 & 0.996 & 1.28(13) & 1.06(7) \\
			$^{228}\!$Ac & 3$^+$ & (0, 0.3) & 0.303 & 0.278 & 0.440(44) & 0.37(13) \\
			\hline 
			\hline 	
		\end{tabular*}		
		\begin{tablenotes}
			\item[a] Extract $g_{K\nu}$ and $g_{K\pi}$ using Eq.~\ref{eq_g_K} with theoretical $\langle s_z \rangle$ values. Here $g_s^\text{eff} = 0.6~g_s^\text{free}$ was applied.
			\item[b] Extract $g_{K\nu}$ and $g_{K\pi}$ values using the experimental values of $g$-factors of neighboring isotopes from Refs.~\cite{Lynch, Verstraelen} and this work and Eq.~\ref{eq_g_odd_A}.			  
		\end{tablenotes}
		\label{table_odd}
	\end{threeparttable}
\end{table*}

\subsection{Quadrupole moments}\label{Q}

Under the strong coupling assumption in this deformation region, the intrinsic quadrupole moment $Q_\text{intr}$ can be deduced from the spectroscopic quadrupole moment $Q_s$ in the case of axial symmetry, using the relation:
	\begin{equation}\label{eq_Q_intr}
		\begin{aligned}
			Q_\text{intr} = \dfrac{(2I+3)(I+1)}{3K^2-I(I+1)}Q_s \\
		\end{aligned}
	\end{equation}
    
\begin{figure}[!h]
	\centering
	\includegraphics[width=0.9\textwidth]{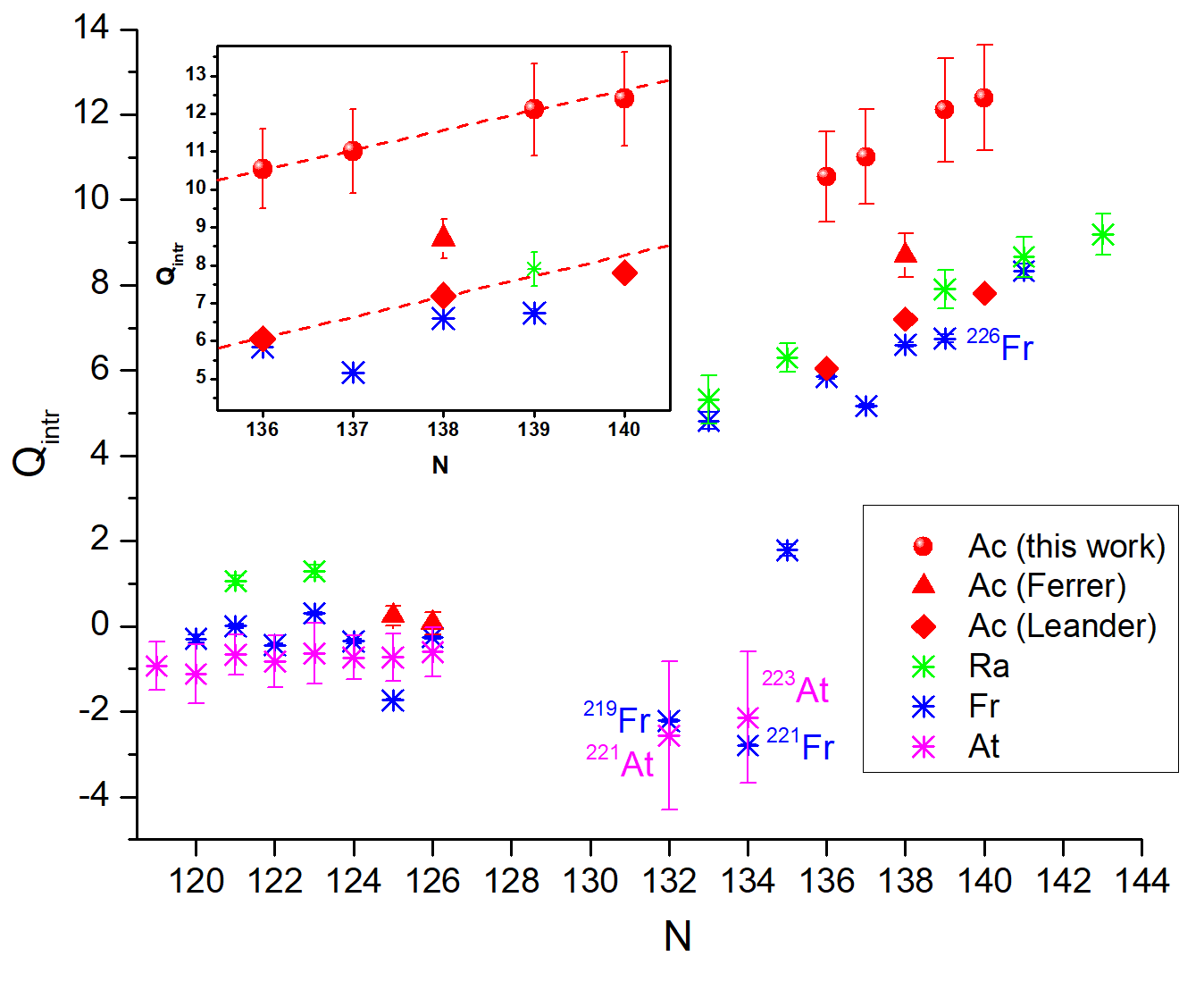}   
	\caption{The intrinsic quadrupole moment $Q_\text{intr}$ of Ac compared to neighboring 
    At~\cite{Barzakh, Cubiss}, Fr~\cite{Budincevic, Coc, Groote}, and Ra~\cite{Wendt, Lynch} isotopes. Several sudden drops of $Q_\text{intr}$ of Fr isotopes on the neutron-rich side are attributed to Coriolis mixing~\cite{Ekstrom, Groote}. Ac data include the experimental data of this work, previously measured $^{227}\!$Ac by Ferrer et al.~\cite{Ferrer} and theoretical values calculated by Leander et al.~\cite{Leander}. The inlet plot shows the details of data in $N$ = 136-140 region.}
	\label{fig:Q}
\end{figure}

$K$~=~$I$ is assumed for well-deformed nuclei \cite{Lynch}, except in special cases such as $^{226}$Fr, where $K$~=~0 and $I$~=~1 are used, as in Ref.~\cite{Ekstrom}. The calculated intrinsic quadrupole moments of Ac isotopes are plotted in Fig.~\ref{fig:Q}, with a comparison with neighboring Ra, Fr, and At isotopes. Several transitional Fr isotopes, $^{219}$Fr, $^{221}$Fr, and $^{226}$Fr (mentioned above), exhibit $K\!\!<I$ due to Coriolis mixing~\cite{Groote}, which manifests as a sudden drop in the spectroscopic quadrupole moment to a negative value, deviating from the strongly positive trend observed in other isotopes at $N\!\!>$ 130 in the isotopic chain. It is also evident in the transitional nuclei $^{221}\!$At and $^{223}\!$At. 

An excellent agreement between the $Q_\text{intr}$ values extracted from $Q_s$ and that from the probabilities of electric quadrupole $\gamma$-transition, $B(E2; 0_1^+\rightarrow2_1^+)$, was observed on \mbox{neutron-rich} $_{88}$Ra, $_{90}$Th, and $_{92}$U (presented in Fig.~8 of Ref.~\cite{Lynch}). Similar to these isotopes with even $Z$, the $Q_\text{intr}$ of Ac ($Z$=89), both our experimental data and Leander's theoretical result, also show a linear increase with $N$, with the identical slope (Fig.~\ref{fig:Q}). However, Leander's theoretical result~\cite{Leander} aligns well with the experimental data of $_{88}$Ra and $_{87}$Fr (Fig.~\ref{fig:Q}), while the data of this work exhibits a systematic positive offset. The experimental data for $^{227}\!$Ac, experimentally measured by Ferrer et al.~\cite{Ferrer}, also shows a positive offset, though to a lesser extent. 

Unlike the absolute values of $Q_\text{intr}$, its slope with $N$ was eliminated from the large theoretical uncertainty in the atomic factors. The experimental uncertainties of $Q_\text{intr}$ (see Tab.~\ref{table_u}) are rather small and practically invisible in Fig.~\ref{fig:Q} in contrast to the $\pm$10$\%$ theoretical uncertainty. If ignoring the systematic offset, the slope of $Q_\text{intr}$ of this work matches that of Leander's theoretical result and Ra/Fr data. A subtle change in the slope of $Q_\text{intr}$ is observed to occur at $N$~=~139-140 in this work. This may be physically linked to the restoration of reflection symmetry after crossing the OES boundary at $N$~=~138. Further experiments on the more neutron-rich isotopes are needed to confirm this observation and gain deeper insight.

\begin{figure}[!h]
	\centering
	\includegraphics[width=1\textwidth]{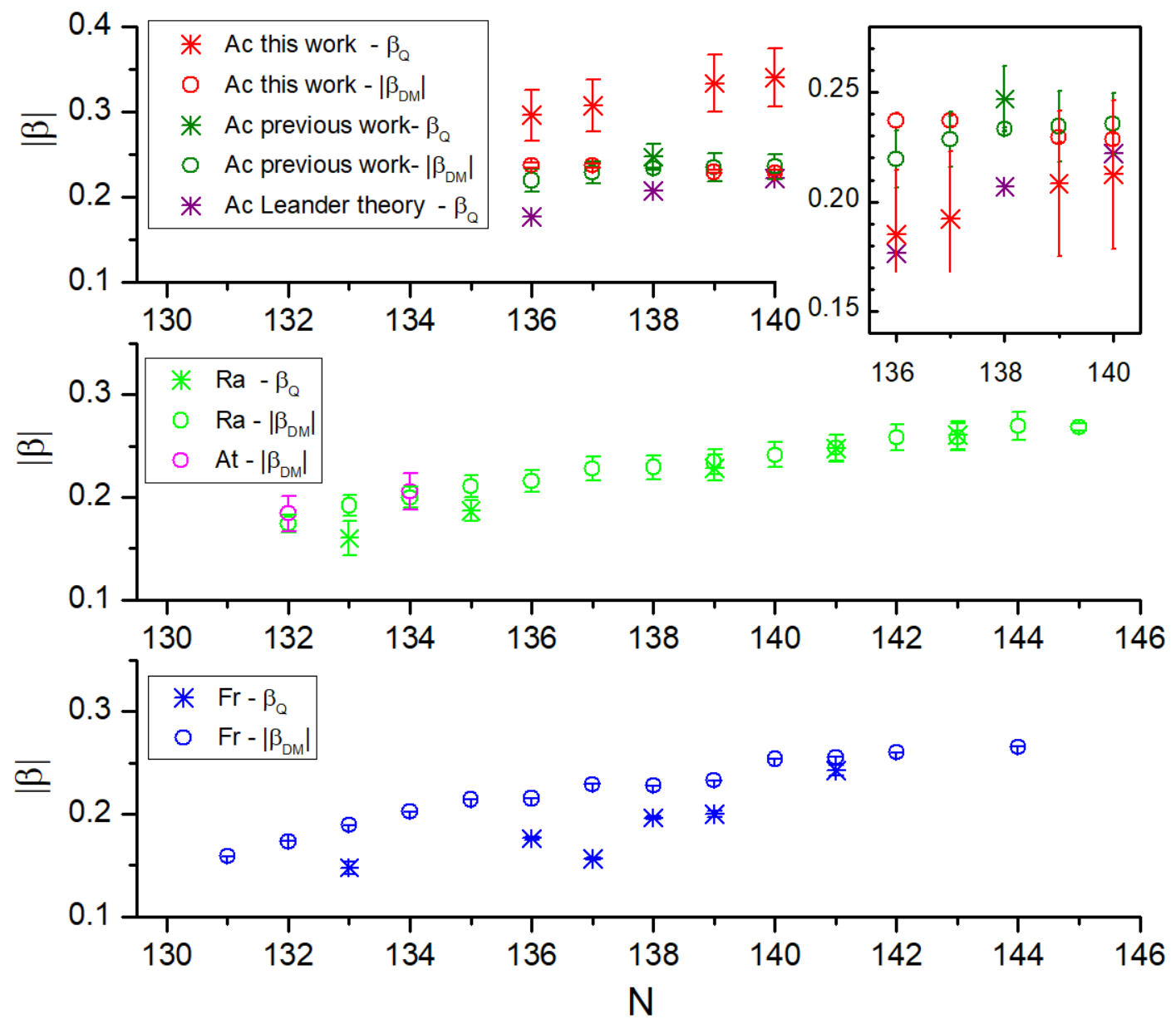}   
	\caption{Comparison of the deformation parameters $\beta_Q$ and $|\beta_{DM}|$ extracted from experimental quadrupole moments and from the changes in mean-square radii of Ac and neighboring Fr~\cite{Budincevic, Coc, Groote}, Ra~\cite{Wendt, Lynch, Wansbeek}, and At~\cite{Barzakh, Cubiss} isotopes. Ac data include previous measurements~\cite{Ferrer, Verstraelen} and theoretical values~\cite{Leander}. The top-right inset plot shows the case if we scale our data using the theoretical $\beta_Q$ values of Leander et al. \cite{Leander} at $N$~=~136 and 140. }
	\label{fig:beta}
\end{figure}	

The nuclear quadrupole deformation parameter $\beta_2$ can be obtained from the $Q_\text{intr}$ using:
\begin{equation}\label{eq_Q_2}
	\begin{aligned}
		Q_\text{intr}\approx\dfrac{5e}{\sqrt{5\pi}}Z\langle r^2 \rangle_{sphr} \langle\beta_2\rangle(1+\dfrac{1}{7}\sqrt{\dfrac{20}{\pi}}\langle\beta_2\rangle) \\
	\end{aligned}
\end{equation}
Here $\langle r^2 \rangle_{sphr}$ is the mean-square charge radius of the nucleus, assuming spherical shape with the same volume, and using the droplet model~\cite{Berdichevsky}. The $\langle\beta_2\rangle$ value extracted from $Q_\text{intr}$ using Eq.~\ref{eq_Q_2} is denoted as $\beta_Q$. 

Meanwhile, the mean-square charge radius can also be expressed as the function of quadrupole deformation parameter $\langle\beta_2^2\rangle$ :
\begin{equation}\label{eq_r^2}
	\begin{aligned}
		\langle r^2 \rangle\approx \langle r^2 \rangle _{sphr} (1+\dfrac{5}{4\pi}\langle\beta_2^2\rangle) \\
	\end{aligned}
\end{equation}

Here higher multipole deformation terms are neglected since the quadrupole term typically dominates the collective shape change. We extracted $\sqrt{\langle\beta_2^2\rangle}$ using Eq.~\ref{eq_r^2} from measured mean-square radii, and denoted them as $|\beta_{DM}|$ to distinguish from $\beta_Q$ obtained from quadrupole moments. Both $|\beta_{DM}|$ and $\beta_Q$ of Ac, Fr, Ra, and At are plotted in Fig.~\ref{fig:beta}. $\beta_Q$ indicates the static quadrupole deformation and is sensitive to the sign of the deformation. It is positive for prolate deformation and negative for oblate deformation. $|\beta_{DM}|$ represents the mean-square quadrupole deformation. The difference between them can be attributed to a dynamic component given by $\beta_{dyn}^2 = \langle \beta_2^2 \rangle- \langle\beta_2\rangle^2$~=~$|\beta_{DM}|^2$-$\beta_Q^2$. This dynamic contribution makes the mean-square charge radius larger than what would be expected from the static deformation alone. It reveals the level of \textquotedblleft$\beta$-softness" of the (axial) deformation in the absence of octupole or triaxial deformation \cite{Moore}. 

In both the Fr and Ra chains, the dynamic deformation decreases with increasing $N$, which implies a progressive hardening of the core towards a rigid system, i.e., a gradual transformation from transitional nuclei to well-deformed nuclei. In the Ra chain, $|\beta_{DM}|$ $\simeq$ $\beta_Q$ is reached at a smaller $N$ than for Fr isotopes. This agrees with the observation in Ref.~\cite{Lynch} that the slope of $Q_{intr}$ with $N$ decreases from $_{88}$Ra through $_{90}$Th to $_{92}$U. This infers the \textquotedblleft$\beta$-softness" hardens fast with increasing proton number. Following these observed rules, the Ac chain should reach the rigidity earlier than the Fr chain, namely before $N$=141.

Since there were no experimental measurements on $Q_s$ of Ac before this work, except one data point at $^{227}\!$Ac, the tendency of this \textquotedblleft$\beta$-softness" could not be revealed. If comparing $|\beta_{DM}|$ deduced from Verstraelen's work~\cite{Verstraelen} and $\beta_Q$ deduced from Leander's theoretical calculation, $|\beta_{DM}|$~$\simeq$~$\beta_Q$ reaches around $^{227}\!$Ac. The $\beta_Q$ values obtained in this work exhibit a systematic positive offset away from previous works, as discussed before. This offset causes $\beta_Q$ higher than the $|\beta_{DM}|$, with $|\beta_{DM}|$ either extracted from this work and Verstraelen's. This contradicts the \textquotedblleft$\beta$-softness" theory and the observations supported it. The lack of $^{227}\!$Ac data in this work makes it impossible to scale our data to previous experiments. However, If we assume the systematic error originates from this work and scale our data using the theoretical $\beta_Q$ values of Leander et al. \cite{Leander} at $N$~=~136 and 140, a similar asymptotic approach of $\beta_Q$ toward $|\beta_{DM}|$ with increasing $N$ presents, as shown in the top-right inset plot in Fig.~\ref{fig:beta}.

\section{Conclusion}
Fast-beam collinear laser spectroscopy on neutron-rich Ac$^+$ was conducted to determine the isotope shifts, HF constant $A$ and $B$ of $^{225, 226, 228, 229}\!$Ac$^+$ at the ionic transition $7s$$^2$ $^1$S$_0$ $\rightarrow$ $6d7p$ $^1$P$_1$. Combined with theoretical atomic factors from multi-configuration Dirac-Hartree-Fock (MCDHF) calculations, the changes in mean-square charge radii, magnetic dipole and spectroscopic electric quadrupole moments of these Ac isotopes were obtained. The improvement in precision allowed us to locate the boundary of the inverse OES in the Ac isotope chain in the region of octupole deformation. The measured spectroscopic quadrupole moments indicate a smooth, gradual increase in deformation, similar to the trends observed in neighboring Fr, Ra, and Th isotopes. A subtle change in the slope of $Q_\text{intr}$ with $N$ was observed at $N$~=~139-140. A more comprehensive understanding requires high-resolution spectroscopy of Ac isotopes at $N\!>$ 140. Additionally, to address the systematic discrepancies between our data and previous studies, which hinder the determination of the rigidity point of dynamic deformation on the Ac chain, CLS spectroscopy of $^{227}\!$Ac is necessary to serve as a reference for comparison.

\begin{acknowledgments}
The experimental work is funded by TRIUMF which receives federal funding via a contribution agreement with the National Research Council of Canada and through Natural Sciences and Engineering Research Council of Canada (NSERC) and individual Discovery Grants RGP-IN-2021-02993 to R.~Li, RGP-IN-2024-659 to J.~Lassen, and RGP-IN-2018-04030 to M. Stachura.
The work on the atomic-structure calculations was supported by the National Natural Science Foundation of China~(Grant Nos. 12474250 and 11874090). 
\end{acknowledgments}


\end{document}